\documentclass[12pt]{article}

\usepackage{vmargin}   
\setmarginsrb{3cm}{2cm}{3cm}{2cm}{1cm}{1cm}{1cm}{1cm}   
\usepackage{amsmath}   
\usepackage{amssymb}  
\usepackage[dvips]{graphicx}  
\usepackage{rotating}   
\usepackage{psfrag} 
\usepackage{fancyhdr}

\newcommand{\Hp}{{\cal H}_{phys}}

\newcommand{\be}{\nopagebreak[3]\begin{equation}}
\newcommand{\ee}{\end{equation}}
\newcommand{\ba}{\nopagebreak[3]\begin{eqnarray}}
\newcommand{\ea}{\end{eqnarray}}

\author{Karim Noui\thanks{noui@gravity.psu.edu} \hspace{0.2cm}and Alejandro Perez \thanks{perez@gravity.psu.edu}\\ \\
{\it Center for Gravitational Physics and Geometry }\\
{\it Pennsylvania State University} \\
{\it University Park, PA 16802, USA}}
\title{\bf Three dimensional loop quantum gravity: coupling to point particles}
\date{\today} 
 
\begin{document} 

\maketitle

\begin{abstract} 
{ We consider the coupling between three dimensional gravity with zero
cosmological constant and massive spinning point particles. First, we study
the classical canonical analysis of the coupled system.  Then, we go to the
Hamiltonian quantization generalizing loop quantum gravity techniques. We give
a complete description of the kinematical Hilbert space of the coupled
system. Finally, we define the physical Hilbert space of the system of
self-gravitating massive spinning point particles using Rovelli's generalized
projection operator which can be represented as a sum over spin foam amplitudes. In
addition we provide an explicit expression of the (physical) distance operator
between two particles which is defined as a Dirac observable.}
\end{abstract} 

\newpage
 
\section*{1. Motivations}
Loop Quantum Gravity provides a promising proposal
of quantum theory of gravity \cite{Ashtekar2, carlobook}. In
contrast to usual quantum field theories, it is a background
independent, non-peturbative approach to the problem of unification of
general relativity and quantum mechanics. 
{Loop quantum gravity is based on the canonical formulation of
general relativity expressed in terms of connection variables. The
quantization program can be presented in terms of two (technically)
distinguishable steps: in the first step the constraints governing
general covariant dynamics are represented as quantum operators in a
Hilbert space of kinematical states. The second step consists of
characterizing the physical Hilbert space of solutions of the quantum
constraints.}  {The first step is well understood. The kinematical
Hilbert space is essentially spanned by polymer-like excitations known
as spin-networks. Moreover, they provide a definition of the notion of
quantum geometry of space as they correspond (in a suitable sense) to
eigenstates of geometric operators. The quantization of the scalar
constraint (governing the quantum dynamics) has been achieved although
due to certain ambiguities in the regularization several proposals
(mathematically equally consistent) have been put forward. The
ambiguity is expected to be resolved by the restrictions that would
arise in trying to reproduce the semiclassical limit of the theory.
In order to achieve this one would need to resolve the second step
mentioned above, i.e.,to characterize the physical Hilbert space for a
given quantization of the scalar constraint.  This second step remains
to a large degree open in loop quantum gravity.}
{The spin foam approach provides a systematic tool to describe the
dynamics aiming at the construction of a mathematically well defined
notion of path integral for loop quantum gravity \cite{Perez}.}  {
For a novel proposal to tackle dynamics in loop quantum gravity which
also leads to a spin foam representation see \cite{phoenix}}

{Despite the difficulties associated to dynamical issues, important
physical questions have been successfully addressed by the theory.}  
In particular, loop quantum gravity provides a
microscopical derivation of black whole entropy \cite{Rovelli3}
and predicts that space is discrete at the Planck scale
\cite{Rovelli2}. {More recently, novel insights about 
the nature of the classical big-bang singularity and the physics of
initial conditions have been achieved in the context of loop quantum
cosmology \cite{Bojowald}.}
The prediction of discreteness of space at the Planck scale is certainly the most
important result of loop quantum gravity; playing a crucial role in the other
physical predictions of the theory. 
{Although the possibility of direct observation of discreteness seems not plausible at this stage, one must ask  }
the question at the conceptual level, i.e., 
is it possible to imagine a physical process that could
measure, without any ambiguity, geometrical properties of space, as
lengths, areas or volumes? 
Making measurements of geometrical properties of space-time is meaningless
without considering the presence of an observer in the theory
\cite{Rovelli4}. In practice, this means that one has to couple the
gravitational field to a physical field associated to the observer in
order to construct physical geometrical operators or Dirac observables
in the Hamiltonian framework. However, this idea is technically very
complicated to implement in general. {Already at the classical
level the definition of observable quantities is quite involved.
Aside for a few examples that are useful in special cases (e.g. the
area of black hole isolated horizons \cite{Ashtekar2}) the construction of
Dirac observables in the quantum theory remains quite open.}


Three dimensional gravity offers a good framework to test these
ideas. In the absence of external field, it is a topological field
theory and therefore only admits a finite number of degrees of
freedom. This property makes the classical theory exactly solvable and
different quantization schemes have been proposed in the literature
\cite{Carlip}. In particular, the loop quantum gravity program can be
completely realized in that case: the kinematical Hilbert space is
well understood and, in a companion paper \cite{Noui}, we make use of
spin-foam models to solve explicitely the dynamics. 
Geometrical operators, as length operators, have been studied and
their spectrum have been computed in both Euclidean \cite{Rovelli5}
and Lorentzian regimes \cite{Freidel}. These operators
are not Dirac observables and cannot be, a priori, related to a physical
process. In the spirit of the previous lines, the simplest way to
construct geometrical observables is to couple the gravitational field
with particles and consider distances between these particles. At the
classical level, this has been successfully done using the
Chern-Simons formulation of gravity \cite{Buffenoir}. Distance
observables between particles are given explicitely, they form a
Poisson algebra which naturally exhibits solutions of the classical
dynamical Yang Baxter equation. This very nice algebraic result
generalizes the combinatorial formalism of pure Chern-Simons theory
\cite{BNR} and allows to consider quantizing the coupled system by the 
use of dynamical quantum groups.

Nevertheless, in this article we will consider the loop quantization
of three dimensional gravity coupled to massive and spinning point
particles. First, we construct classical distance observables in the
Dirac sense.  We then proceed to the quantization, find their
spectrum, and compare it with (non physical) distance operators
previously considered. We describe the full dynamics of the
self-gravitating massive spinning particles system in terms of the
spin foam representation. Recently, the inclusion of particles in the
Ponzano-Regge model has been considered in a rather different context
\cite{Freidel2}.

{The paper is organized as follows.}  In the next section, we
present the classical coupling between point particles and the three
dimensional gravitational field: we adapt the results of
\cite{Buffenoir} in the case of Euclidean gravity with zero
cosmological constant and we give an explicit expression for the
distance between particles which are Dirac observables. In the third
section, we quantize the coupled system \`a la loop quantum gravity:
we first give a complete description of the kinematical Hilbert space,
then we construct the physical Hilbert space.
To describe the kinematical Hilbert space, we extend the
notion of spin-network and we define generalized spin-network states
which include particles.
To construct the physical Hilbert space, we generalize the construction presented in
the companion paper \cite{Noui}: we propose a regularization for the
projector on physical states and we obtain a description of the
physical scalar product in terms of the spin foam representation. The obtained
spin foam model generalizes the Ponzano-Regge model in the sense that
it includes massive and spinning point particles { and, as in
\cite{Noui}, it is not defined on a background triangulation as all
regulators are removed in the definition of the physical inner
product.}

\section*{2. Classical formulation of the theory}

Leaving out the problem of degenerate metrics, it is well known
\cite{Witten} that three dimensional gravity is equivalent to a
Chern-Simons theory whose gauge group $G$ depends on the signature of
the metric and on the sign of the cosmological constant. Furthermore,
the coupling between a point particle and the gravitational field can
be formulated as a minimal gauge coupling \cite{Buffenoir,Sousa}.

This section aims at using this formulation to make a precise classical
analysis of the coupling between particles and Euclidean gravity
without cosmological constant. For that purpose, we will first present
a non-usual formulation to describe the dynamics of a free Euclidean
relativistic particle in such a way that we could immediately apply,
in a second step, gauge theory methods to describe the coupled system.

\subsection*{2.1. Free relativistic particle} 

Our formulation is based on the fact that we
can identify the degrees of freedom of a massive spinning
relativistic particle evolving in the three dimensional Euclidean
space ${\mathbb E}^3$ with an element of the (universal covering of
the) isometry group $G=ISU(2)$ of ${\mathbb E}^3$ (see for example
\cite{Sousa}). Indeed, any element $X \in G$ can be decomposed, by
definition, as the semi-direct product $X=(\Lambda,q)$ of a rotation
$\Lambda \in SU(2)$ and a translation $q$. Then, we can obviously
identify the element $q$ with the position of a particle in ${\mathbb
E}^3$ and we will see latter that $\Lambda$ allows to construct a
``spin-vector'' associated to the particle \cite{Sousa}. In the
sequel, we make extensive use of the vectorial representation
of the group $G$ where $\Lambda$ is a $3\times 3$ rotation matrix and
$q$ is a three-vector of ${\mathbb E}^3$. Notation and properties of
the Euclidean group $G=ISU(2)$ and its Lie algebra $\mathfrak
g=isu(2)$ are exposed in the appendix.

	    \subsubsection*{2.1.1. Algebraic description}
Given a point particle of mass $m$ and spin $s$, its degrees of freedom are
completely determined by an element $X=(\Lambda,q)$ of the euclidean group $G$
and its dynamics between times $t_1$ and $t_2$ is defined by the following
first order action \cite{Sousa}:
\begin{eqnarray}\label{freeparticleaction}
S_p[X] \; = \; \int_{t_1}^{t_2} dt \; <\chi(m,s)\; , \; X^{-1} \frac{dX}{dt}>
\;\; \text{where} \;\;\; \chi(m,s)=mJ_0 + sP_0 \in \mathfrak g \;.
\end{eqnarray}
The application $<,>: \mathfrak g \times \mathfrak g \rightarrow \mathbb C$ is
a non-degenerate invariant bilinear form on $\mathfrak g$ (appendix). We will
introduce the notation ${\cal C} \simeq G$ to denote the configuration space
of the point particle. In that framework, Euclidean transformations (analog of
Poincar\'e transformations) of the particle degrees of freedom are defined by
the following map:
\begin{eqnarray}\label{Poincaretransformations}
{\cal C} \; \times \; G \; \longrightarrow {\cal C} \;\;\;\;\; , \;\;\;\;\;\; (g,X) \; \longmapsto \; gX \;.
\end{eqnarray}
Applying these transformations in the vectorial representation of $G$, one can
in particular recover the well-known Euclidean transformations of the position
$q$ of the particle.

Before going into a precise canonical analysis, let us show in the Lagrangian
framework that $S_p$ reproduces the motion of a free massive spinning
particle on $\mathbb E^3$. For that purpose, we compute the variation of the
action $\delta S_p$ with respect to $X$ and we see immediately that the
variational problem is well defined when the quantity $<\chi(m,s) \; , \;
X^{-1} \delta X>$ vanishes on the boundaries $t=t_1$ and $t=t_2$. Actually,
this condition fixes half of the degrees of freedom of the particle at initial
and final times and, in the particular case of a purely massive particle
($s=0$), it is equivalent to the condition $\delta q=0$, i.e. the position of
the particle is fixed on the boundaries. With these rather standard boundary
conditions the equations of motion are
\begin{eqnarray}\label{equationsofmotionforfreeparticle}
[\chi(m,s) \; , \; X^{-1}\frac{dX}{dt}] \; = \; 0 \;\; \Longleftrightarrow
\;\; X^{-1} \frac{dX}{dt} \; = \; \alpha J_0 + \beta P_0 \;,
\end{eqnarray}
where $[,]$ is the commutator in the Lie algebra. 

In the last equation, $\alpha$ and $\beta$ are, \`a priori, arbitrary
functions of time. The equivalence (\ref{equationsofmotionforfreeparticle}) is
trivial and general solutions of the equations of motion read $X(t)=X(t_1)
\exp \int_{t_1}^{t_2} (\alpha J_0 + \beta P_0)$ where $X(t_1)$ is an element
of $G$ defined by initial conditions. In order to clarify the relation between
these solutions and the motion of a massive spinning particle in $\mathbb
E^3$, let us write them explicitely in the vectorial representation:
\begin{eqnarray}
\Lambda(t) \; = \; \Lambda(t_1) \exp \int_{t_1}^{t_2}\!\! dt \; \alpha J_0
\;\;\;\; \text{and} \;\;\;\; q(t) \; = \; q(t_1) + T(t)\Lambda(t)P_0 \;.
\end{eqnarray}
We have introduced the notation $T(t)=\int_{t_1}^{t}\beta dt$. As a result,
the dynamics of the particle coordinates takes the form $q^i(t) = q^i(t_1) +
\frac{1}{m} p^i T(t)$ where we have defined $p^i = <J_i , \Lambda P_0> = m
<\Lambda^{-1}J_i \Lambda, P_0>$. Thus we obtain 
the geodesic equations of the metric space ${\mathbb E}^3$. Notice that
$p^i$ are constants of motion satisfying the constraint $p^2 = p^i p^i =m^2$ and
therefore are interpreted as the components of the $3$-momentum of the
particle. Besides, the function $T(t)$ is viewed as a clock\footnote{The clock
$T$ is an arbitrary function but should be a monotonic function of time
parameter in order to describe causal trajectories. Actually, it is possible
to find a derivative gauge fixing compatible with the variational problem and
such that $T(t)$ is monotonic (cf 2.1.3).} associated to the particle which
measures time evolution. Following \cite{Sousa}, the spin vector $\sigma$ of
the particle is proportional to the momentum, i.e. $\sigma^i=\frac{s}{m}p^i$
and, therefore, its dynamics is trivial. Note that the momentum components
(and also the spin) are expressed in a given frame, called {\it internal
frame} in the sequel.

This brief Lagrangian analysis makes a crystal-clear contact between the
dynamics induced by the action (\ref{freeparticleaction}) and the motion of a
standard free particle in $\mathbb E^3$. Furthermore, it is immediate to see
that the dynamics of a free particle in the three dimensional Minkowski space
$\mathbb M^3$ can be formulated in the same way but replacing the gauge group
$G$ by the Poincar\'e group $ISO(2,1)$ \cite{Sousa}. Similar formulation
exists to describe the motion of a particle in the three dimensional de Sitter
space $dS^3$ \cite{Buffenoir} where the gauge group is now replaced by the
isometry group of $dS^3$ and we can extend this formulation to the Anti-de
Sitter case.

This non-usual formulation will appear soon very interesting to describe the
coupling with a gravitational field and then to quantize the obtained
system. 

           \subsubsection*{2.1.2. Canonical analysis}
This section is devoted to the canonical analysis of the free particle
action. Note that we will present the analysis in a formulation such that one
could easily generalize to the Lorentzian, deSitter or anti-deSitter cases for
example. The reader not interested into the details can have the main ideas of
this analysis in the footnote \footnote{The variable $X \in ISU(2)$ contains
all the degrees of freedom of the particle: the position $q^i$ and three
angles labeling $\Lambda \in SU(2)$ which carry the information about the
direction of the momentum of the particle---as we will show below one of the three angles
is a gauge degree of freedom while the other two label a point on the sphere of
directions of $p_i$. From the properties of the
Killing form (see appendix), it is immediate to show that the action can be
rewritten as follows:
\begin{eqnarray}
S_p[q,\Lambda] \; = \; \int_{t_1}^{t_2} dt \left( mn_i \frac{dq^i}{dt}
-\frac{s}{2} \text{Tr}(J_0 \Lambda^{-1} \frac{d\Lambda}{dt})\right)
\;. \nonumber
\end{eqnarray}
In this expression, $n_i$ is a unit vector defined by $n^iJ_i = \Lambda J_0
\Lambda^{-1}$ and $Tr$ is the trace in the fundamental representation of
$SU(2)$.

This action clearly defines a constrained system whose constraints reads
\cite{Sousa}:
\begin{eqnarray}
\Phi_i \; \equiv \; p_i - mn_i \; \simeq \; 0 \;\;\;\; \text{and} \;\;\;\;
\Psi_i \; \equiv \; \sigma_i - sn_i \; \simeq \; 0 \;, \nonumber
\end{eqnarray}
where $p_i$ is the momentum (conjugated to position $q^i$) and $\sigma_i$ is
the spin-vector whose expression in terms of the phase space variables is
given in \cite{Sousa}. In a sense, one can think of the vector-spin components
as variables ``conjugated'' to the angles labeling $\Lambda$. It will be
convenient to consider the total angular momentum $j_i = \lambda_i + \sigma_i$
where the orbital momentum is defined by $\lambda = q \wedge p$.

From the set of six constraints, one can extract two first class constraints
which read:
\begin{eqnarray}
C_p \; \equiv \; p_i p^i - m^2 \; \simeq \; 0 \;\;\;\; \text{and} \;\;\;\; C_j
\; \equiv \; p_i j^i - ms \; \simeq \; 0 \;.\nonumber
\end{eqnarray}
The first constraint is clearly the usual mass shell condition of the
relativistic particle and therefore generates reparametrizations along the
trajectory of the particle. The second one generates rotations of the frame
which conserve the momentum $p^i$ (its orbit corresponds to the third angle
mentioned above). It can be also viewed as a definition of the spin $s$ of
the particle. The analysis of the second class constraints leads to the Dirac
bracket whose expression is given in \cite{Sousa} or equivalently by the
expression (\ref{freeparticleDiracbracket}) in the core of the text.}at the
bottom of this page which draws its inspiration from \cite{Sousa}.

To a general study of the Hamiltonian analysis, it will be convenient first not to distinguish the position and spin degrees of freedom of the particle. Therefore, we consider a parametrization $z=(z^I)_{I=1,\cdots,6} \in \mathbb R^6$ of the group $G$ and we will denote by $X(z)$ the element of $G$ parametrized by $z$. 

We associate to each parameter $z^I$ the canonically conjugated momentum $\pi_I$ such that the Poisson brackets are given by $\{\pi_I,z^J\}=\delta^J_I$. Using the notations introduced in the appendix, it will be convenient to define the 
matrix $P=\varrho_I \xi^I$ where we have considered the variables $\varrho_I = \pi_J \frac{\partial f^J}{\partial \alpha^I}(\alpha=0,z)$. In terms of these new variables, the Poisson brackets read:
\begin{eqnarray}
\{X_1,X_2\}=0 \;\; , \;\; \{P_1,X_2\}=-t_{12} X_2 \;\;\; \text{and} \;\;\; \{P_1,P_2\}=[t_{12},P_1]\;.
\end{eqnarray}
In these equations, we have used the standard tensorial notation and $t \in \mathfrak g^{\otimes 2}$ denotes the Casimir tensor associated the the bilinear form $<,>$.

The free particle action (\ref{freeparticleaction}) clearly defines a constrained system and the constraints analysis is done along the same lines as in the de Sitter case \cite{Buffenoir}. In particular, conjugated momenta are fixed in terms of the coordinates by a set of six primary constraints which can be expressed in the following matricial form:
\begin{eqnarray}\label{primaryconstraints}
C[X,P] \; \equiv \; X^{-1}PX + \chi(m,s) \; \simeq \; 0 \;.
\end{eqnarray}
Once the primary constraints identified, we introduce the Lagrange multiplier $\mu \in C^{\infty}(\mathbb R, \mathfrak g)$ in order to define the total Hamiltonian by $H_{tot}=<\mu,C[X,P]>$ for the canonical Hamiltonian of the theory is identically zero. Using the immediate identity $\pi_I \dot{z^I}=-<P,\dot{M}M^{-1}>$, we see that the free particle action (\ref{freeparticleaction}) is equivalent to the following first order action:
\begin{eqnarray}\label{totalaction}
S_{tot}[X,P,\mu] \; = \; -\int_{t_1}^{t_2}dt \left( <P,\frac{dM}{dt}M^{-1}> - H_{tot}[X,P,\mu]\right)\;.
\end{eqnarray}
The total Hamiltonian $H_{tot}$ is necessary to define temporal evolution of the dynamical variables and we show that conservation of primary constraints under time evolution does not introduce new constraints. However, the Lagrange multiplier $\mu$ is restricted to be an element of the Cartan subalgebra, i.e. $\mu = \mu_P P_0 + \mu_J J_0$. As a result, among the six constraints (\ref{primaryconstraints}), $C_P=<P_0,C[X,P]>$ and $C_J=<J_0,C[X,P]>$ are first class. On the other hand, the functions $C_I=<\xi_I,C[X,P]>$ where $\xi_I \in (P_i,J_i)_{i=1,2}$ form a set of second class constraints whose Dirac matrix is given by:
\begin{eqnarray}\label{Diracmatrix}
\{C_I,C_J\} \; = \; \Delta_{IJ} \; \equiv \; <\chi(\tilde{m},\tilde{s}),[\xi_I,\xi_J]> \; .
\end{eqnarray}
We have introduced the notations $\tilde{m}=m+C_P$ and $\tilde{s}=s+C_J$. Note that it is crucial to invert the Dirac matrix strongly in order to obtain a Dirac bracket which satisfies, in particular, the Jacobi identity. This is the reason why we don't solve explicitely the first class constraints in the expression (\ref{Diracmatrix}) of the Dirac matrix. Taking this remark into account, we compute the inverse Dirac matrix and we obtain:
\begin{eqnarray}
(\Delta^{-1})^{IJ} \; = \; <\kappa(\tilde{m},\tilde{s}),[\xi^I,\xi^J]> \;\;\;\; \text{where} \;\;\;\; \kappa(m,s) \; =\; \frac{-mJ_0 + sP_0}{m^2+s^2}\;.
\end{eqnarray}
As a consequence, it is possible to compute Dirac bracket between any two functions $\varphi$ and $\psi$ defined on the phase space. In particular, Poisson brackets between matrix elements of $P$ and any function on the phase space are not modified by the Dirac procedure for $P$ commutes with the second class constraints. On the other hand, Poisson brackets between matrix elements of $X$ are replaced by the following Dirac bracket:
\begin{eqnarray}\label{freeparticleDiracbracket}
\{X_1,X_2\}_D \; = \; X_1 X_2 r_{12}(\tilde{m},\tilde{s}) \;.  
\end{eqnarray}
 As a consequence, the Dirac bracket appears to be quadratic and introduces the $r$-matrix $r:\mathbb R^2 -\{(0,0)\}\rightarrow \mathfrak g \otimes \mathfrak g$ defined, for any non zero couple $(m,s)$, by the expression $r_{12}(m,s)=[\kappa_2(m,s),t_{12}]$. The antisymmetry property of the Dirac bracket implies that the symmetric part of $r$ vanishes, i.e. $r_{12} + r_{21}=0$. Moreover, the Jacobi identity is equivalent to the fact that $r$ is a solution of the classical dynamical Yang-Baxter (CDYB) equation associated to the Lie-algebra $\mathfrak g$ \cite{Babelon,Buffenoir}. This result is particularly interesting and underlines an unexpected link between the dynamics of a free relativistic particle and solutions of the CDYB equation. This relation could be the starting point of a quantization program for the relativistic particle involving quantum groups but we will adopt a more conservative point of view in this article.

From the Dirac bracket (\ref{freeparticleDiracbracket}), one can establish the equations of motion of the dynamical variables in the Hamiltonian framework and one obtains:
\begin{eqnarray}\label{freeparticulehamiltonianequationsofmotion}
\frac{dX}{dt} \; = \; \{X,H_{tot}\}_D \; = \; X\mu \;\;\; \text{and} \;\;\; \frac{dP}{dt} \; = \; \{P,H_{tot}\}_D \; = \; 0 \;.
\end{eqnarray}
As one could expect for a free particle, the conjugated momentum $P$ is a constant of motion. If one decomposes it as follows $P=p^iJ_i + j^iP_i$, then $(p^i)$ and $(j^i)$ are respectively the momentum and the total angular momentum of the particle whose expressions are explicitely given by the formulae:
\begin{eqnarray}\label{momentumandangularmomentum}
p^iJ_i \; = \; -m \Lambda \; J_0 \; \Lambda^{-1} \;\;\; \text{and} \;\;\; j^iP_i \; = \; +m\Lambda J_0 \Lambda^{-1} \; q \; - \; s\Lambda \; P_0 \;.
\end{eqnarray}
As we could expect, the total angular momentum is the sum of the angular momentum $\lambda=q\wedge p$ and the spin-vector $\sigma=\frac{s}{m}p$. Using the two invariant bilinear forms on $\mathfrak g$ (appendix), one can determine the relativistic invariants of the free particle as follows:
\begin{eqnarray}
<P,P> \; = \; 2p \cdot j \; = \; 2ms \;\;\;\;\; \text{and} \;\;\;\;\; (P,P) \; = \; p \cdot p \; = m^2 \;.
\end{eqnarray}
The dot $\cdot$ denotes the usual scalar product on the three dimensional Euclidean space $\mathbb E^3$. As for the equation of motion for $X$ (\ref{freeparticulehamiltonianequationsofmotion}), it is the same as the one obtained in the Lagrangian framework (\ref{equationsofmotionforfreeparticle}) when one replaces $\alpha$ and $\beta$ respectively by $\mu_J$ and $\mu_P$. As a result, general solutions of the equations of motion read:
\begin{eqnarray}\label{solutionoffreeparticleequationsofmotion}
X(t) \; = \; X(t_1) \exp \int_{t_1}^t du \mu(u) \;\;\;\; \text{and} \;\;\;\; P(t) \; = \; X(t_1) \chi(m,s) X(t_1)^{-1} \;. 
\end{eqnarray}
In the vectorial representation, it is clear that these solutions reflect the dynamics of a free relativistic particle in $\mathbb E^3$. At this level, $\mu$ appears to be any function of the time with values in the Cartan subalgebra of $\mathfrak g$ and thus there is no restriction on the evolution of the particle position along its straightline trajectory. For instance, it is possible to find $\mu$ such that the particle oscillates along its trajectory and then its motion violates causality. In fact, the freedom to choose any particular evolution for the particle comes from the invariance of the system under time reparametrizations. As a consequence, one can fix this problem by an appropriate gauge fixing \cite{Buffenoir, Teteilboim}. For that purpose, it is necessary to make a study of the symmetries of the system.

          \subsubsection*{2.1.3. Symmetries and time reparametrization}
Among different types of symmetries, we distinguish gauge symmetries (local) and Noether symmetries (global). In the Hamiltonian formulation, the former are generated by first class constraints whereas constants of motion generate the latter. 

In our formulation of the relativistic particle, the generator of infinitesimal gauge transformations is given by $G[\varepsilon]=<\varepsilon,C[X,P]>$ where $\varepsilon$ is an infinitesimal element of the Cartan subalgebra of $\mathfrak g$ and gauge symmetries on the dynamical variables read:
\begin{eqnarray}\label{freeparticlegaugetransformations}
\delta_\varepsilon X \; = \; \{G[\varepsilon],X\} \; = \; -X\varepsilon \;\;\;\; \text{and} \;\;\;\; \delta_\varepsilon P \; = \; \{G[\varepsilon],P\} \; = \; 0 \;. 
\end{eqnarray}
When we evaluate the previous Poisson brackets, we assume that the infinitesimal parameter $\varepsilon$ does not depend on the dynamical variables. However, it appears that the action (\ref{freeparticleaction}) transforms as $\delta_\varepsilon S = <\varepsilon(t_2) - \varepsilon(t_1),\chi(m,s)>$ under these transformations and therefore is generically not let invariant. As a result, the invariance of the action is obtained if one imposes the vanishing of $<\varepsilon, \chi(m,s)>$ on the boundaries. In that case, the total action (\ref{totalaction}) is invariant under the transformations (\ref{freeparticlegaugetransformations}) if and only if the Lagrange multiplier $\mu$ transforms as $\delta_\varepsilon \mu = - \dot{\varepsilon}$. At this point, it is very interesting to note that, if one defines the function $M \in C^{\infty}([t_1,t_2],G)$ by the expression
\begin{eqnarray}\label{dynamicalfunction}
M(t)\; = \; \exp-\int_{t_1}^t du \mu(u) \;, 
\end{eqnarray}
then the quantity $X(t)M(t)$ is invariant under gauge symmetries and therefore is an observable of the system. In fact, from solutions of the equations of motion (\ref{solutionoffreeparticleequationsofmotion}), we see immediately that $X(t)M(t)=X(t_1)$ is fixed by initial conditions. In the sequel, $M(t)$ will be referred as the dynamical function of the particle for it contains all the dynamical information.

To clarify the relation between these gauge transformations and usual symmetries of the free relativistic particle, let us remark that $\delta_\varepsilon X = -\lambda \dot{X} -\lambda(X\mu - \dot{X})$ when $\varepsilon=\lambda \mu$ and $\lambda$ is an arbitrary function of $t$. As a result, one first notes that gauge symmetries are equivalent, up to the equations of motion, to reparametrization symmetries of the free particle action (\ref{freeparticleaction}), i.e. $t \mapsto t - \lambda(t)$. However, gauge transformations contain more than the usual reparametrization symmetry of the particle trajectory for they are generated by two different first class constraints which can, in fact, be rewritten as $C_P=\frac{1}{m} p^2  - m$ and $C_J = \frac{1}{m}p\cdot j -s$. Thus, one can show that the former explicitely implements time reparametrizations of the particle position $q$ whereas the latter generates rotations of the internal frame which let invariant the momentum $p$.

We can ask the question of a gauge fixing compatible with the boundary condition satisfying by the parameter $\varepsilon$. The derivative gauge fixation $<\dot{\mu},\chi(m,s)>=0$ is a good candidate. To be more precise, this is a partial gauge fixing and the remaining gauge freedom can be fixed using a canonical gauge. In the purely massive case, it implies the fixation of the component $\mu_P$ of the Lagrange multiplier and therefore the clock $T(t)$ associated to the particle is a constant (so monotonic) function.

\medskip

Before going to the coupled system, it is important to note that the free particle action (\ref{freeparticleaction}) is invariant under the global transformations (\ref{Poincaretransformations}). This symmetry reflects the freedom to choose the frame (origin and axes) of the Euclidean space ${\mathbb E}^3$. The generators of these infinitesimal symmetries are given by the constants of motion $p^i, j^i$ and we have:
\begin{eqnarray}\label{infinitesimalPoincaresymmetries}
\{p^i,X\}_D \; = \; -P_i X \;\;\;\;\; \text{and} \;\;\;\;\; \{j^i,X\}_D \; = \;-J_i X \; . 
\end{eqnarray}
Moreover, $p^i$ and $j^i$ commute with the constraints and therefore are observables of the theory. It is easy to show that their Poisson algebra satisfies the Euclidean algebra, i.e.:
\begin{eqnarray}
\{p^i,p^k\}_D \; = \; 0 \;\; , \;\; \{j^i,j^k\}_D \; = \; \epsilon^{ik}{}_l \; j^l \;\; \text{and} \;\; \{p^i,j^k\}_D \; = \; \epsilon^{ik}{}_l \; p^l \;.
\end{eqnarray}
The coupling to the gravitational field consists precisely of gauging the global symmetries (\ref{infinitesimalPoincaresymmetries}) using, as usual, a $G$-connection $\cal A$. However, a naive minimal gauge coupling introduces ambiguities and therefore it is necessary to consider a regularization. This point was completely understood in the de Sitter case \cite{Buffenoir} and next section is devoted to recall the basic ideas in our case without entering into the details.

         \subsection*{2.2. Coupling to the gravitational field}
The fact that there is no gravitational interaction between point particles in the three dimensional case allows one to study the coupling to each particle with the gravitational field independently. For this reason, we will endeavor to present the case of one particle coupled to gravity; the generalization to an arbitrary number of particles will be immediate. We will consider a three dimensional manifold $\cal M$ whose topology is given by ${\cal M}=\Sigma \times [t_1,t_2]$ and we will denote by $x_p$ the trajectory of the particle.

	    \subsubsection*{2.2.1. Minimal gauge coupling and regularization}
Up to boundary terms, the first order action of three dimensional Euclidean gravity defined on the manifold $\cal M$ can be formulated as a Chern-Simons action whose gauge group is $G=ISU(2)$:
\begin{eqnarray}\label{ChernSimonsaction}
S_{CS}[{\cal A}] \; = \; \int_{\cal M} d^3x \; \epsilon^{\mu \nu \rho} \left( <{\cal A}_\mu,\partial_\nu {\cal A}_\rho> + \frac{1}{3} <{\cal A}_\mu,[{\cal A}_\nu,{\cal A}_\rho]> \right) \;.
\end{eqnarray}
The connection ${\cal A}={\cal A}_\mu dx^\mu$ is related to the spin connection $A^i = A_\mu^i dx^\mu$ and the triad $e^i = e_\mu^i dx^\mu$ by the expression ${\cal A}_\mu = e_\mu^i P_i + A_\mu^i J_i$. This action defines a gauge theory and infinitesimal gauge transformations of the connection are given by:
\begin{eqnarray}\label{connectiongaugetransformations}
\forall \; \xi \in C^{\infty}({\cal M},\mathfrak g) \;,\;\;\; \delta_\xi {\cal A}_\mu \; = \; {\cal D}_\mu \xi \; \equiv \; \partial_\mu \xi +[{\cal A}_\mu,\xi] \;.
\end{eqnarray}
We can use the Chern-Simons connection to gauge the global symmetries of the free particle (\ref{infinitesimalPoincaresymmetries}) such that the coupled action reads:
\begin{eqnarray}\label{naivecoupling}
S_c[{\cal A},X] \; = \; \int_{t_1}^{t_2} dt <\chi(m,s),X^{-1}\frac{dX}{dt} + X^{-1} {\cal A}_t X> \;.
\end{eqnarray}
The integral is defined along the trajectory of the particle and is, by construction, invariant under local transformations (\ref{infinitesimalPoincaresymmetries}) when the connection transforms as (\ref{connectiongaugetransformations}). Finally, the dynamics of the whole coupled system (particle and gravity) is governed by the action:
\begin{eqnarray}
S[{\cal A},X] \; = \; S_{CS}[{\cal A}]+ S_c[{\cal A},X] \;.
\end{eqnarray}
Equations of motion are obtained by varying this action with respect to the dynamical variables. In particular, the time component of the connection ${\cal A}_t$ appears to be a Lagrange multiplier which imposes the following constraints on the curvature ${\cal F}$ of the Chern-Simons connection $\cal A$:
\begin{eqnarray}\label{conicalsingularity}
\forall \; x \; \in \; {\Sigma}  , \;\;  \epsilon^{ab}{\cal F}_{ab}(x) \; = \; P\; \delta(x-x_p) \;  \Longleftrightarrow \;  \left \{
\begin{array}{rl}
\epsilon^{ab}T^i_{ab}(x) & =  j^i \; \delta(x-x_p) \\
\epsilon^{ab}F^i_{ab}(x) & =  p^i \; \delta(x-x_p) 
\end{array}
\right.
\end{eqnarray}
When we decompose l.h.s. equations (\ref{conicalsingularity}) into its translational and rotational parts, we obtain well-known constraints (r.h.s.) on the torsion tensor $T[e,A]$ and the curvature tensor $F[A]$. As we could expect, we find that the connection is singular at the location of the particle: this is the signature of the well-known conical singularity induced by a particle in three dimensional gravity \cite{Deser}. However, the equations (\ref{conicalsingularity}) can only fix the conjugacy class of the holonomy of the connection along a loop surrounding the particle unless we specify a given origin on the surface $\Sigma$. This equation is then ill-defined and it is necessary to reformulate the coupling (\ref{naivecoupling}) in order to regularize it. Different schemes already exist \cite{Buffenoir, Matschull} and we will adopt the one developed in the article \cite{Buffenoir} that the reader is invited to read if she is interested by the details.

\medskip

The basic idea consists of replacing the point particle by a small loop $\ell$ whose time evolution draws a cylinder ${\cal B}=\ell \times [t_1,t_2]$ in the space-time ${\cal M}$. In such a picture, the coupling is defined as an integral over ${\cal B}$ which trivially generalizes the action (\ref{naivecoupling}) and we have:
\begin{eqnarray}\label{actionfortheboundary}
S_c^{reg}[{\cal A},X] \; = \; \frac{1}{2\pi} \int_{\cal B}dt \; d\varphi <\chi(m,s),X^{-1}\frac{dX}{dt} + X^{-1}{\cal A}_t X> \;.
\end{eqnarray}
Note that $\varphi \in [0,2\pi[$ is the angular variable parametrizing the loop $\ell$ and the origin $\varphi=0$ is arbitrary. In this action, $X$ still depends only on $t$ and we see that the coupling is, in fact, defined with the mean value of ${\cal A}_t$ along the loop $\ell$.

The dynamics of the gauge field ${\cal A}$ is, \`a priori, described by a Chern-Simons action (\ref{ChernSimonsaction}). However, because of the regularization, the space-time admits a boundary $\cal B$ which breaks the gauge invariance. To restore the invariance, first we have to ask the question of what type of symmetry we would like at the boundary. We will naturally extend the gauge transformations (\ref{connectiongaugetransformations}) to the case where the manifold $\cal M$ admits a boundary but we will require that the infinitesimal gauge parameter $\xi$ evaluated on any point of $\cal B$ does not depend on the angle $\varphi$ which does not have any physical meaning. Finally, to have a theory invariant under these transformations, the Chern-Simons action have to be supplemented with the following boundary term:
\begin{eqnarray}\label{boundaryterm}
S_{\cal B}[{\cal A}] \; = \; \int_{\cal B}dt \; d\varphi <{\cal A}_t,{\cal A}_\varphi> \;.
\end{eqnarray}
It is trivial to verify that the resulting action $S_{CS}[{\cal A}]+S_{\cal B}[{\cal A}]$ is invariant under the required gauge transformations. Then, the regularized action for the coupled system is given by:
\begin{eqnarray}\label{regularizedcouplingsystem}
S^{reg}[{\cal A},X] \; = \; S_{CS}[{\cal A}] \; + \; S_{\cal B}[{\cal A}] \; + \; S_c^{reg}[{\cal A},X]\;.
\end{eqnarray}
The generalization to an arbitrary number $N$ of particles $(\wp_i)_{i=1,\cdots,N}$ is straightforward: one associates a boundary ${\cal B}_i=\ell_i \times [t_1,t_2]$ to any particle whose dynamics is governed by the action (\ref{actionfortheboundary}) and one adds as many boundary terms (\ref{boundaryterm}) as there are particles to fix the problem of the symmetry breaking. The dynamical variable of the particle $\wp_i$ of mass $m_i$ and spin $s_i$ will be denoted by $X_i=(\Lambda_i,q_i) \in G$.

	    \subsubsection*{2.2.2. Hamiltonian analysis: symmetries and observables}
This section aims to present the main results of the canonical analysis of the coupled action (\ref{regularizedcouplingsystem}). The reader interested in the technical aspects is invited to read the article \cite{Buffenoir}. In particular, the constraint (\ref{conicalsingularity}) is regularized and is replaced by the following constraint defined for any $v \in C^{\infty}(\Sigma,\mathfrak g)$:
\begin{eqnarray}\label{regularizedconstraint}
\Phi(v) \; \equiv \; \int_\Sigma d^2x \; \epsilon^{ab}<v,{\cal F}_{ab}> + 2 \int_\ell d\varphi <v,{\cal A}_\varphi + \frac{1}{2\pi}X\chi(m,s)X^{-1}> \;.
\end{eqnarray}
As a consequence, the connection is flat on the surface $\Sigma$ but its holonomy $H_\ell$ around the loop $\ell$ is given by $H_\ell=X \exp(-\chi(m,s))X^{-1}$. Note that there is no more ambiguity and the singularity has been removed: the connection ${\cal A}$ is now defined everywhere on $\Sigma$.

When $v$ is constant along $\ell$, then the constraint (\ref{regularizedconstraint}) is first class and therefore generates gauge transformations. Besides, the canonical analysis shows that the constraints (\ref{primaryconstraints}) remain and $G(\varepsilon) = <\varepsilon,C>$ are still first class constraints when $\varepsilon$ is an element of the Cartan subalgebra of $\mathfrak g$. Let us denote by $\alpha =(\varepsilon,v)$ the set of gauge transformations parameters and the gauge transformations generator is defined by $\Gamma(\alpha)=G(\varepsilon) + \Phi(v)$. Thus, the symmetries generated by $\Gamma(\alpha)$ on the variables $X$ and on the spatial components of the connection $({\cal A}_a)_{a=1,2}$ read:
\begin{eqnarray}\label{gaugetransformationsofcouplingsystem}
\delta_\alpha X \; = \; -X\varepsilon - v(\ell)X \;\;\;\;\; \text{and} \;\;\;\;\; \delta_\alpha {\cal A}_a \; = \; {\cal D}_av \;
\end{eqnarray}
where $v(\ell)$ denotes the evaluation of the function $v$ on the loop $\ell$. These results are trivially extended to the case where there is an arbitrary number $N$ of particles and we will denote by $\varepsilon_i$ the infinitesimal parameter of the reparametrization symmetry of the particle $\wp_i$ and $v(\ell_i)$ the evaluation of $v$ on the loop $\ell_i$.

\medskip

In the absence of particles, the theory reduces to a Chern-Simons theory associated to the group $G$ and the physical phase space is known as the moduli space of flat $G$-connections. Classical observables are defined in terms of spin-networks whose Poisson algebra is well-known \cite{Goldman}. Different attempts to quantize the theory has been developed the last fifteen years but the combinatorial quantization method is indisputably the most powerful (see \cite{BNR} and references therein). This quantization scheme uses quantum groups as a cental tool. 

\begin{figure}
\psfrag{phii}{$\varphi_i$}
\psfrag{phij}{$\varphi_j$}
\psfrag{U}{$\gamma$}
\psfrag{pi}{$\wp_i$}
\psfrag{pj}{$\wp_j$}
\psfrag{0}{$0$}
\centering
\includegraphics[scale=1]{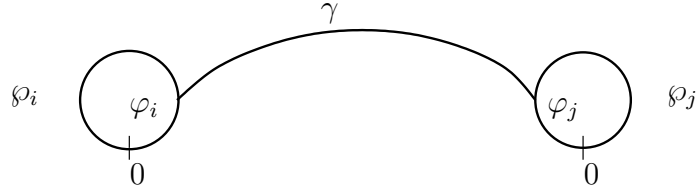}
\caption{\small Example of a generalized spin-network observable.}
\label{figdistance}
\end{figure}

The combinatorial description of Chern-Simons theory can be generalized in the presence of point particles \cite{Buffenoir}. In particular, we can define new observables which capture the dynamics of the particle and contain, in particular, the relative positions between particles. To understand this point, let us define the following function:
\begin{eqnarray}\label{configurationoperator}
{\cal O}_\gamma[{\cal A},X_i,X_j] \; \equiv \; M_i(t)^{-1}X_i(t)^{-1} U_\gamma[{\cal A}] X_j(t)M_j(t) \;.
\end{eqnarray}
In this definition, $U_\gamma[{\cal A}]$ denotes the holonomy of the connection along the curve $\gamma$ whose starting point and end point are respectively the points of coordinates $\varphi_i \in \ell_i$ and $\varphi_j \in \ell_j$ (figure \ref{figdistance}); $M_i \in C^{\infty}([t_1,t_2],G)$ is the dynamical function of the particle $\wp_i$ (\ref{dynamicalfunction}). 
As the connection $\cal A$ is flat, ${\cal O}_\gamma$ depends only on the homotopy class of the path $\gamma$. Moreover, it is clearly invariant under the gauge transformations (\ref{gaugetransformationsofcouplingsystem}) and therefore is an observable. Similarly to the free particle case, equations of motion imply that $X_i(t) M_i(t) = X_i(t_1)$ represents the initial configuration of the particle $\wp_i$: this is the reason of the invariance of the observable (\ref{configurationoperator}) under reparametrizations of the particles. 

The observable ${\cal O}_\gamma[{\cal A},X_i,X_j]$ can be interpreted as the (initial) configuration variable of the particle $\wp_j$ measured in the rest frame of the particle $\wp_i$. To be more explicit, we shall write ${\cal O}_\gamma[{\cal A},X_i,X_j]$ in the vectorial representation and we obtain:
\begin{eqnarray}\label{Configurationinthefundamentalrepresentation}
\left(
\begin{array}{cc}
\Lambda_i(t_1)^{-1} h_\gamma[A] \Lambda_j(t_1) & \Lambda_i(t_1)^{-1}(h_\gamma[A]q_j(t_1) + q_{\gamma}[e,A] - q_i(t_1)) \\
0 & 1 \\
\end{array}
\right).
\end{eqnarray}
$h_\gamma[A] \equiv P\exp \int_{\gamma} A$ denotes the holonomy of the spin-connection $A$ along the path $\gamma$ and we have introduced the notation $q_\gamma[e,A] \equiv \int_{\gamma} h_{\gamma<x}[A] e_a(x) dx^a$ where the path $\gamma<x$ (resp. $\gamma>x$) is the part of $\gamma$ which ends to (resp. starts from) the point $x \in \gamma$. Therefore, the translational part ${q}_\gamma(i,j)$ of ${\cal O}_\gamma$ (\ref{Configurationinthefundamentalrepresentation}) clearly measures the position  of the particle $\wp_j$ in the rest frame of $\wp_i$ \cite{Buffenoir}. Then, the relative distance ${\cal D}_\gamma(i,j)$ is given by the formula:
\begin{eqnarray}\label{distanceoperator}
 {\cal D}_\gamma^2(i,j) \; = \; {q}_\gamma(i,j)^{\dagger} \; {q}_\gamma(i,j) \;.
\end{eqnarray}
The rotational part $\Lambda_\gamma(i,j)$ of ${\cal O}_\gamma$ represents the (rotation) transformation which sends any observer in the rest frame of the first particle to any observer in the rest frame of the second one. In particular,  it allows to compute the momentum components $p^k_\gamma(i,j)$ of the particle $\wp_j$ in the rest frame of $\wp_i$ as follows:
\begin{eqnarray}
p^k_\gamma(i,j) \; = \; m_j <P_k \; , \; \Lambda_{\gamma}(i,j) J_0 \Lambda_{\gamma}(i,j)^{-1}> \;.
\end{eqnarray}
We will denote by $p_{rest}^k(i)$ the components of the momentum of $\wp_i$ in its rest frame.

The distance operator (\ref{distanceoperator}) is a gauge invariant function which depends only on the homotopy class of $\gamma$. Furthermore, when the particles are purely massive ($s_i=0$), ${\cal D}_\gamma(i,j)$ does not depend on the choice of the starting point $\varphi_i \in \ell_i$ and the end point $\varphi_j \in \ell_j$. When particles possess non trivial spins, the local geometry around one given particle $\wp_i$ says that two different points of $\ell_i$ are identified but at different time. As a result, one does not expect the distance ${\cal D}_\gamma(i,j)$ to be independent of $\varphi_i$ and $\varphi_j$, which is indeed the case.
 
Let us remark that we deal with the particles reparametrization invariance by considering initial configuration variables in the definition of the observables (\ref{configurationoperator}) which makes use of the Lagrange multipliers $\mu_i$. Therefore, these observables do not capture the dynamics of each particle. Actually, it is necessary to consider a gauge fixing to recover the dynamics in the same way as we did in the free particle case (cf 2.1.3). However, it is possible to construct non trivial observables without considering any choice of initial conditions. A first example of such observables is given by the following matrix element:
\begin{eqnarray}\label{boostparameter}
\theta_\gamma(i,j) \; = \; <\stackrel{1}{e}_0 \vert \; \Lambda_1(t)^{-1} h_\gamma[A] \Lambda_2(t) \; \vert \stackrel{1}{e}_0> \;.
\end{eqnarray}
Notations are recalled in the appendix. It is straightforward to show the identity $\theta_\gamma(i,j)=p_\gamma(i,j)\cdot p_{rest}(i)$ and thus (\ref{boostparameter}) is physically interpreted as the {\it Euclidean boost} parameter between the two particles. An other interesting example reads as follows:
\begin{eqnarray}
d_\gamma(i,j) \; = \; \frac{\vert q_\gamma(i,j) \cdot p_\gamma(i,j) \wedge p_{rest}(i)\vert}{\parallel p_\gamma(i,j) \wedge p_{rest}(i) \parallel} \;.
\end{eqnarray}
This expression gives the minimal distance between the trajectories of the two particles, is clearly gauge invariant and does not depend on the choice of an initial space slice.

\medskip

The analysis of second class constraints is done along the same lines as in \cite{Buffenoir}. One can compute the Dirac matrix, inverse it and therefore get the Dirac bracket. It appears that the Dirac bracket between spin-networks which do not touch the boundaries is still given by the Goldman symplectic structure \cite{Goldman}. Moreover, the Dirac bracket between particles degrees of freedom is not modified and is still given by (\ref{freeparticleDiracbracket}). However, the Dirac bracket between functions depending on the boundaries connections is generically very complicated. In the sequel, we will refer to the Dirac phase space ${\cal P}_D$ the set of classical states once the secondary constraints have been implemented. Any point $x \in {\cal P}_D$ is completely determined by a $G$-connection $\cal A$ and a sequence $(X_1, \cdots, X_N)$ of elements of $G$. 

The set of observables (\ref{configurationoperator}) forms a Poisson algebra whose structure is very nice and introduces solutions of the CDYB equation \cite{Buffenoir}. This result motivates a study of the quantization using combinatorial formalism techniques. However, the method is far from being straightforward and needs to be developed to understand first representation theory of non-compact dynamical quantum groups and then to construct unitary irreducible representations of the observables quantum algebra. This program clearly deserves a complete study.

The following aims to explore loop quantum gravity techniques as an interesting alternative of the quantization. In particular, we define the notion of kinematical Hilbert space of the coupled system, we quantize the distance observables and compute their action on particular states. Then, we propose a spin-foam picture to define the physical scalar product generalizing the results obtained in the companion paper \cite{Noui}.

\section*{3. Loop quantization}
Loop quantum gravity is an attempt to describe a background independent quantization of general relativity (see \cite{Ashtekar2} for a review). This program clearly distinguishes two different steps: the description of the kinematical Hilbert space in terms of spin-networks and the analysis of the dynamics using spin-foam models \cite{Perez}. 

This section aims to adapt loop quantization techniques in the case of three dimensional gravity coupled to point particles. For that purpose, we decompose the first class constraints $\Phi$ (\ref{regularizedconstraint}) into its torsion part $\Phi_T$ and its curvature part $\Phi_C$ defined by:
\begin{eqnarray}\label{separationofconstraints}
\Phi_T(v) \; \equiv \; \Phi(v^i J_i) \;\;\;\; \text{and} \;\;\;\; \Phi_C(v) \; \equiv \; \Phi(v^iP_i)
\end{eqnarray}
for any element $v \in C^{\infty}(\Sigma,\mathbb R^3)$ which is constant on the boundary $\ell$. These constraints respectively generate the rotational and the translational gauge transformations (\ref{gaugetransformationsofcouplingsystem}) on the dynamical variables. In fact, they are regularized versions of the ill-defined constraints (r.h.s. \ref{conicalsingularity}). 

The strategy will consist of quantizing the theory before implementing the first class constraints. The kinematical Hilbert space ${\cal H}_0$ will be defined as the set of states satisfying all first class constraints but not the curvature constraint $\Phi_C$ endowed with a suitable scalar product. Then, we will define the physical Hilbert space ${\cal H}_{phy}$ from the kinematical one by introducing a projector which implements the constraint $\Phi_C$. Finally, we will make use of this projector to define a spin-foam model describing dynamics of massive spinning particles coupled to the gravitational field.

         \subsection*{3.1. Kinematical Hilbert space}
The first step consists of quantizing the Dirac phase space ${\cal P}_D$ endowed with its Dirac symplectic structure. The Dirac bracket between particles degrees of freedom is given by the quadratic bracket (\ref{freeparticleDiracbracket}) and the Dirac bracket between geometrical degrees of freedom (not leaving on the boundaries) can be written as follows:
\begin{eqnarray}\label{geometricalbracket}
\{e_a^k(x) \; , \; A_b^l(y)\}_D \; = \; \epsilon_{ab} \; \eta^{kl} \; \delta(x-y) \;\;, \; \forall \; x,y \; \in \; \Sigma - \cup_{i} \ell_i \; .
\end{eqnarray}
All the other Dirac brackets vanish but the Dirac bracket between geometrical degrees of freedom leaving on the boundaries of $\Sigma$ which are generically very complicated. However, we won't consider, in this article, this boundary Dirac bracket which, nevertheless, deserves a detailed study we postpone for future investigations \cite{Noui2}. 

As a result, we can quantize independently geometrical and particles degrees of freedom. Loop quantum gravity provides a straightforward quantization of the geometrical part (\ref{geometricalbracket}). The quantization of the particle degrees of freedom is presented in the first subsection. Then, we describe the kinematical Hilbert space of the coupled system. Finally, we compute the action of distance operators on particular kinematical states.

	   \subsubsection*{3.1.1. Quantum relativistic particle}
At the kinematical level, the particle $\wp$ is classically characterized by its position $q$ and an element $\Lambda \in SU(2)$. The system is quantized \`a la Schrodinger and we consider the polarization such that $\Lambda$ is the configuration variable. This choice of polarization is the analog of the momentum polarization in the usual formulation of the free relativistic particle. 

We will denote by $\text{Pol}[\Lambda]$ the vector space of polynomial functions of $\Lambda$ whose a basis is given by matrix elements of unitary representations of $SU(2)$, i.e.:
\begin{eqnarray}\label{polynomialfunctionofLambda}
\text{Pol}[\Lambda] \; = \; \oplus_{i \in \frac{1}{2}\mathbb N} \text{Pol}_i[\Lambda] \;\;\; \text{where} \;\; \text{Pol}_i[\Lambda] \; \equiv \; \{\stackrel{i}{\pi}{}\!\!(\Lambda)^l_n \; \vert \; l,n \in [-i,i]\} \;.
\end{eqnarray}
Notations and basic notions on representation theory of $SU(2)$ are presented in the appendix. We endow this vector space with a Hilbert structure making use of the $SU(2)$ Haar measure $d\mu$ as follows:
\begin{eqnarray}\label{particleHaarmeasure}
\forall \; f,g \; \in \; \text{Pol}[\Lambda] \;\; , \;\;\; <f , g> \equiv \; \int d\mu(\Lambda) \; \overline{f(\Lambda)}g(\Lambda) \;.
\end{eqnarray}
We will denote by ${\cal H}_{0,i}(\wp)$ the Hilbert space obtained by completing $\text{Pol}_i[\Lambda]$ with respect to (\ref{particleHaarmeasure}) and the resulting Hilbert space ${\cal H}_{0}(\wp) \equiv \oplus_i {\cal H}_{0,i}(\wp)$ will be called the free particle kinematical Hilbert space in the sequel. 

At this point, we can define an action of quantum observables on ${\cal H}_{0,i}(\wp)$. For that purpose, let us recall that the quantum algebra of observables is generated by the elements $\hat{p}^k$ and $\hat{j}^k$, $k=0,1,2$, which satisfy the following commutation relations:
\begin{eqnarray}\label{observablesquantumalgebra}
[\hat{p}^k \; , \; \hat{p}^l] \; = \; 0 \;\; , \;\;  [\hat{p}^k \; , \; \hat{j}^l] \; = \; \epsilon^{kl}{}_n \; \hat{p}_n \;\; , \;\;  [\hat{j}^k \; , \; \hat{j}^l] \; = \; \epsilon^{kl}{}_n \; \hat{j}_n \;. 
\end{eqnarray}
From the choice of the polarization, it is natural to think that $\hat{p}^k$ acts by multiplication and $\hat{j}^k$ acts as a derivation in the free particle kinematical Hilbert space. To be more precise, one can check that the following action on ${\cal H}_{0,i}(\wp)$:
\begin{eqnarray}
 \hat{p}^k \; \rhd \; \stackrel{i}{\pi}{}\!\!(\Lambda)^l_n & = & p^k \; \stackrel{i}{\pi}{}\!\!(\Lambda)^l_n \; = \;  -m<P_0,\Lambda^{-1}J_k\Lambda> \; \stackrel{i}{\pi}{}\!\!(\Lambda)^l_n \;,\\ 
 \hat{j}^k \; \rhd \; \stackrel{i}{\pi}{}\!\!(\Lambda)^l_n & = &  \partial^{L}_k \; \stackrel{i}{\pi}{}\!\!(\Lambda)^l_n \; = \;\;  \stackrel{i}{\pi}{}\!\!(J_k)^l_r  \; \stackrel{i}{\pi}{}\!\!(\Lambda)^r_n \;,
\end{eqnarray}
defines a unitary irreducible representation of (\ref{observablesquantumalgebra}). We have introduced the notation $\partial_k^L$ for the left-invariant derivation on the group $SU(2)$. For any $i \in \frac{1}{2}\mathbb N$, the basis (\ref{polynomialfunctionofLambda}) of $\text{Pol}_i[\Lambda]$ diagonalizes the angular momentum whose eigenvalues are obtained from the following actions:
\begin{eqnarray}
\hat{j}_0 \; \rhd \; \stackrel{i}{\pi}{}\!\!(\Lambda)^l_n \; = \; l  \; \stackrel{i}{\pi}{}\!\!(\Lambda)^l_n \;\;\; \text{and} \;\;\; \hat{j}^2 \; \rhd \; \stackrel{i}{\pi}{}\!\!(\Lambda)^l_n \; = \;i(i+1) \; \stackrel{i}{\pi}{}\!\!(\Lambda)^l_n \;.
\end{eqnarray}

To define the physical Hilbert space of the system, a preliminary work consists of quantizing the constraints and computing their action on ${\cal H}_0(\wp)$. The constraint $C_P=\frac{1}{m}p^2 -m$ is quantize without ambiguity for the components of the momentum commute. Thus, we have $\hat{C}_P \equiv \frac{1}{m}\hat{p}^2 - m$ and it is straightforward to see that:
\begin{eqnarray}
\forall \; f \in {\cal H}_0(\wp)\;\;,\;\;\;\; \hat{C}_P \rhd f \; = \; 0. 
\end{eqnarray}
As for the constraint $C_J=\frac{1}{m}p\cdot j + s$, it is also quantize without any ambiguity and a direct calculation shows that the action of $\hat{C}_J \equiv \frac{1}{m}\hat{p}\cdot \hat{j} - s$ on $\text{Pol}[\Lambda]$ reads:
\begin{eqnarray} 
\forall \; i \in \frac{1}{2}\mathbb N \;\; \text{and} \;\; \forall \; l,n \in [-i,i] \;\;,\;\;\;\; \hat{C}_J \rhd \stackrel{i}{\pi}{}\!\!(\Lambda)^l_n \; = \; (n - s) \;  \stackrel{i}{\pi}{}\!\!(\Lambda)^l_n \;.
\end{eqnarray}
As a result, the vector space of physical states is the sub-space of $\text{Pol}[\Lambda]$ (\ref{polynomialfunctionofLambda}) defined by the following direct sum:
\begin{eqnarray}\label{vectorspaceofphysicalstates}
\text{Pol}[\Lambda]^{inv} \; \equiv \; \bigoplus_{i-s \; \in {\mathbb N}} \{\stackrel{i}{\pi}{}\!(\Lambda)^l_{s} \; \vert \; l \; \in \; [-i,i]\}\;.
\end{eqnarray}
As we could expect (in the Euclidean case), we remark immediately that the spin of the particle $s$ must be half-integer. Moreover, the direct sum over $SU(2)$ finite dimensional representations (\ref{vectorspaceofphysicalstates}) is restricted to representations $i$ such that $i-s$ is a positive integer. In fact, this result is a straightforward consequence of the fact that the total angular momentum $j$ is the sum (\ref{momentumandangularmomentum}) of the vector spin $\sigma$ and the orbital momentum $\lambda$ which admits integer eigenvalues in the quantum theory. In the case of a spinless particle ($s=0$), the vector space of physical states (\ref{vectorspaceofphysicalstates}) can be trivially identified to the space of polynomial functions on the two-sphere $S^2 = SU(2)/U(1)$, i.e. the space of spherical harmonic functions. The $SU(2)$ Haar measure naturally endows (\ref{vectorspaceofphysicalstates}) with a Hilbert structure and the Hilbert space ${\cal H}({\wp})$ obtained by completion is precisely the physical Hilbert space of the free relativistic particle.  

The quantization of a system of an arbitrary number $N$ of non-interacting relativistic particles ${\cal P} \equiv \{\wp_k \vert k=1,\cdots,N\}$ is straightforward. The physical Hilbert space of the system ${\cal H}({\cal P})$ consists of the tensor product of the physical Hilbert space of each particle, i.e. ${\cal H}({\cal P}) =\otimes_k {\cal H}({\wp_k})$.

Before going to the coupled system, it is interesting to note that the previous description gives a ``background independent'' description of the dynamics of relativistic particles in the sense that we do not really make use of the metric structure of $\mathbb E^3$. Given a manifold ${\cal M}=\Sigma \times [t_1,t_2]$, the dynamics of each particle $\wp_k$ is characterized by a worldline drawn on $\cal M$. In the Hamiltonian framework, each particle $\wp_k$ is associated to a point $x_k \in \Sigma$ endowed with a mass $m_k$ and a spin $s_k$. As a result, a physical state can be described by a set of points on $\Sigma$ colored by a couple $(i_k,l_k)\in \frac{1}{2}\mathbb N \times [-i_k,i_k]$. Such a description is clearly the analog of spin-network states for quantum gravity. 

At this point, one can ask the question whether it is possible to extract geometric informations (as distance between particles) from such a ``spin-network state''. A priori, the distance between two particles $\wp_k$ and $\wp_n$ is clearly not an observable and therefore its action on the physical Hilbert space ${\cal H}({\cal P})$ is not well-defined. However, the minimal distance $d_{min}(k,n)$ between two worldlines is a gauge invariant quantity (i.e. invariant under reparametrizations) and obviously contains geometric informations; its expression reads:
\begin{eqnarray}
d_{min}(k,n) \; \equiv \; \frac{1}{m_k m_n} \vert (q_k - q_n) \cdot (p_k \wedge p_n) \vert \; = \;  \frac{1}{m_k m_n} \vert j_k \cdot p_n + j_n \cdot p_k \vert \;.
\end{eqnarray}
It is straightforward to quantize this observable, to compute its action on the two-particle state $\stackrel{i_k}{\pi}{}\!(\Lambda_k)^{l_k}_{s_k} \otimes \stackrel{i_n}{\pi}{}\!(\Lambda_n)^{l_n}_{s_n}$ and to notice that it is generically not diagonal. In the case where the particle $\wp_n$ is in rest, i.e. $\Lambda_n=1$, then the action simplifies drastically and is given by:
\begin{eqnarray}
\hat{d}_{min}(k,n) \; \rhd \; \stackrel{i_k}{\pi}{}\!(\Lambda_k)^{l_k}_{s_k} \otimes 1 \; = \; l_k \; \stackrel{i_k}{\pi}{}\!(\Lambda_k)^{l_k}_{s_k} \otimes 1 \;.
\end{eqnarray}
The choice $\Lambda_n=1$ physically corresponds to consider the particle $\wp_n$ as a classical observer who makes measurements on the quantum system. In particular, its minimal distance to the particle $\wp_k$ is quantized and is given by the magnetic number $l_k$ of the particle.

           \subsubsection*{3.1.2. Quantum kinematical states}
This subsection aims to give a construction of the kinematical Hilbert space of the coupled system in analogy with the pure gravity case \cite{Noui}. 

First, let us recall the definition of the space of cylindrical functions $\text{Cyl}(\Sigma)$ on the surface $\Sigma$. Given on oriented graph $\Gamma$ immersed in $\Sigma$, we denote the set of its edges and its vertices respectively by ${\cal E}_\Gamma=\{\gamma_1,\cdots,\gamma_{E_\Gamma}\}$ and ${\cal V}_\Gamma=\{v_1,\cdots,v_{V_\Gamma}\}$. An element $\Psi_{\Gamma,f} \in \text{Cyl}(\Sigma)$ is a complex valued functional of the (generalized) spacelike connection $A$ labeled by a finite graph $\Gamma \in \Sigma$  and a continuous function $f:SU(2)^{\times E_\Gamma} \rightarrow \mathbb C$ and defined by:
\begin{eqnarray}
\Psi_{\Gamma,f}[A] \; \equiv \; f(h_{\gamma_1}[A],\cdots,h_{\gamma_{E_\Gamma}}[A]) \;.
\end{eqnarray}
We recall that $h_\gamma[A]$ denotes the holonomy of the connection along the path $\gamma$. Given two cylindrical functions $\Psi_{\Gamma_1,f_1}$ and $\Psi_{\Gamma_2,f_2}$ and the minimal graph $\Upsilon$ containing $\Gamma_1$ and $\Gamma_2$ as subgraphs, then one can trivially extend these two cylindrical functions to cylindrical functions with support on $\Upsilon$ \cite{Gaul}; we will respectively denote by $f_{12}$ and $f_{21}$ the trivial extensions of $f_1$ and $f_2$ on $\Upsilon$. As a result, one can easily endow $\text{Cyl}(\Sigma)$ with a Hilbert structure using the $SU(2)$ Haar measure $d\mu$ as follows:
\begin{eqnarray}\label{innerproductonCyl}
 <\Psi_{\Gamma_1,f_1} , \Psi_{\Gamma_2,f_2}> \; = \; \int \prod_{i=1}^{E_\Upsilon} d\mu(h_i)\; \overline{f_{12}(h_1,\cdots,h_{E_\Upsilon})} \; f_{21}(h_1,\cdots,h_{E_\Upsilon}) \;.
\end{eqnarray}
The completion of $\text{Cyl}(\Sigma)$ under this inner product defines the auxiliary Hilbert space ${\cal H}_{aux}(\Sigma)$. This Hilbert space provides a quantization of the classical Poisson algebra (\ref{geometricalbracket}) in the case where $\Sigma$ has no boundary. To be more precise, it is a quantization of the algebra of holonomies of the connection along any loop on $\Sigma$ and the triad. The former acts by multiplication and the latter as a derivative operator whose action is formally given by the following differential operator:
\begin{eqnarray}
\hat{e}^k_a(x) \; \equiv \; -i\hbar \;  \epsilon_{ab}\; \eta^{kl}\frac{\delta}{\delta A_b^l(x)}\;.
\end{eqnarray}
Our concern here is not to enter into the details of the definition of its action on any cylindrical function which has been extensively studied in the literature (see \cite{Freidel} and references therein). Nevertheless, it will be useful for our purpose to recall its action on a given holonomy $h_\gamma[A]$ which reads:
\begin{eqnarray}\label{derivationofholonomy}
\hat{e}^k_a(x) \; x[\zeta]^a \; \rhd \; h_\gamma[A] \; = \; s(x,\zeta,\gamma) \; h_{\gamma<x}[A] \; J_k \; h_{\gamma>x}[A] \;. 
\end{eqnarray}
In this expression, we have denoted $x[\zeta]=x[\zeta]^a\partial_a$ the tangent vector to a curve $\zeta$ at $x$; the quantity $s(x,\zeta,\gamma)=0$ if $x$ is not a point of $\gamma$; otherwise it is given by the determinant $s(x,\zeta,\gamma)= \epsilon_{ab}x[\zeta]^a x[\gamma]^b$. As a result, it takes value on the set $\{-1,0,1\}$ depending on the relative orientation of the curves $\gamma$ and $\zeta$ at the point $x$. 

 The kinematical Hilbert space of pure gravity ${\cal H}_0(\Sigma)$ is the subspace of gauge invariant functions of ${\cal H}_{aux}(\Sigma)$. Gauge invariant spin-network states provide ${\cal H}_0(\Sigma)$ with an orthonormal basis. The reader is invited to look at the review paper \cite{Ashtekar2} to have a precise derivation of the kinematical Hilbert space in the four dimensional case.

\medskip

The next step consists of constructing the kinematical Hilbert space of the
coupled system. For that purpose, we generalize the notion of space of
cylindrical functions to the following vector space:
\begin{eqnarray}\label{spaceofgeneralizedcylindricalfunctions}
\text{Cyl}(\Sigma,{\cal P}) \; \equiv \; \text{Cyl}(\Sigma) \otimes \text{Pol}[\Lambda_1]^{inv} \otimes \cdots \otimes \text{Pol}[\Lambda_N]^{inv} \;.
\end{eqnarray}
The inner products on $\text{Cyl}(\Sigma)$ (\ref{innerproductonCyl}) and on
$\text{Pol}[\Lambda]$ (\ref{particleHaarmeasure}) naturally endow the space of
generalized cylindrical functions $\text{Cyl}(\Sigma,{\cal P})$ with a Hilbert
structure. The resulting Hilbert space ${\cal H}_{aux}(\Sigma,\cal P)$ will be
called the generalized auxiliary Hilbert space.

\medskip

The generalized Gauss constraint $\Phi_T$ (\ref{separationofconstraints})
defines $SU(2)$ gauge transformations on the variables of the coupled
system. In particular, finite gauge transformations of the holonomy
$h_\gamma[A]$ of the spin-connection $A$ along an oriented path $\gamma$ and
of the matrices $\Lambda_i \in SU(2)$ read:
\begin{eqnarray}\label{SU(2)gaugetransformations}
h_\gamma[\omega] \; \longmapsto \; g(s_\gamma) h_\gamma[A] g(t_\gamma)^{-1}
\;\;\;\; \text{and} \;\;\;\; \Lambda_i \; \longmapsto \; g(\ell_i) \Lambda_i
\;.
\end{eqnarray}
In these expressions, $g \in C^{\infty}(\Sigma, SU(2))$ whose values $g(\ell_i)$ on each circle $\ell_i$ is constant; $s_\gamma$ and $t_\gamma$ are respectively the source and target points of $\gamma$.

By duality, the action (\ref{SU(2)gaugetransformations}) induces an action (a co-action to be more precise) on ${\cal H}_{aux}(\Sigma,\cal P)$ and the kinematical Hilbert space of the coupled system ${\cal H}_0(\Sigma,\cal P)$ is defined by the Hilbert subspace of gauge invariant functions of ${\cal H}_{aux}(\Sigma,\cal P)$.
\begin{figure}
\psfrag{i1}{$i_1$}
\psfrag{i2}{$i_2$}
\psfrag{i3}{$i_3$}
\psfrag{ell}{$\ell$}
\psfrag{iota}{$\iota_\ell$}
\psfrag{j}{$j_\ell$}
\centering
\includegraphics[scale=0.5]{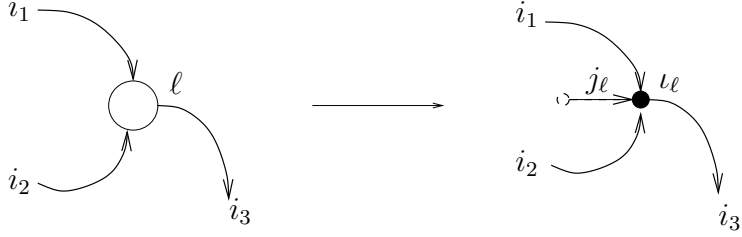}
\caption{\small Example of a boundary vertex. The boundary $\ell$ is associated to a representation $j_\ell$ virtually represented by an extra link oriented toward the boundary; the magnetic number associated to the departure point is fixed to the spin $s$ of the particle. The boundary is also colored with an intertwiner $\iota_\ell$ from in-coming representations (containing the virtual one) to out-going representations.}
\label{boundaryspinnetwork}
\end{figure}

In order to provide ${\cal H}_0(\Sigma,\cal P)$ with an orthonormal basis, we need to introduce the notion of generalized spin-network. For that purpose, we consider an oriented graph $\Gamma$ immersed in the surface with boundaries $\Sigma$. Each edge $\gamma \in {\cal E}_\Gamma$ is colored as usual by assigning to it a finite dimensional irreducible representation $j_{\gamma}$ of $SU(2)$. Any vertex $v \in {\cal V}_\Gamma$ such that $v$ does not belong to the boundaries $\cup_i \ell_i$ is associated to an $SU(2)$-intertwiner $\iota_v$ from in-coming representations to out-going representations. So far, everything is similar to the usual definition of spin-network but the novelty is that the boundaries of $\Sigma$ are also colored. Indeed, we associate to each boundary $\ell_i$ a couple $(j_{\ell_i},\iota_{\ell_i})$ of a finite dimensional irreducible representation $j_{\ell_i}$ and a intertwiner $\iota_{\ell_i}$ of $SU(2)$ in the way presented in the picture (\ref{boundaryspinnetwork}).

A generalized spin-network is defined by $S \equiv \{\Gamma,(j_\gamma)_{\gamma \in {\cal E}_\Gamma},(\iota_v)_{v \in {\cal V}_\Gamma},(j_{\ell_i},\iota_{\ell_i})_i\}$. To each generalized spin-network $S$, we associate an element $\Psi_S \in {\cal H}_0(\Sigma,\cal P)$ (called a generalized spin-network state) as follows:
\begin{eqnarray}
\Psi_S[A;(\Lambda_i)_i] \; \equiv \; (\bigotimes_{i=1}^N \iota_{\ell_i})(\bigotimes_{v\in{\cal V}_{\Gamma}} \iota_v)(\bigotimes_{i=1}^N \stackrel{j_{\ell_i}}{\pi}{}\!(\Lambda_i)_{s_i})(\bigotimes_{\gamma \in {\cal E}_\Gamma} \stackrel{j_\gamma}{\pi}{}\!(h_\gamma[A])) \;.
\end{eqnarray}

This definition is a straightforward generalization of the usual one. Note
that there is one free magnetic number associated to each particle link
$\gamma_{\ell_i}$ which contains the information of the spin $s_i$ of the
particle $\wp_i$. It is clear that $\Psi_S[A;(\Lambda_i)_i]$ is invariant
under $SU(2)$-gauge transformations (\ref{SU(2)gaugetransformations}) and
therefore is an element of ${\cal H}_0(\Sigma,\cal P)$. Moreover, the set of
generalized spin-network states form a complete orthonormal basis in the
generalized kinematical Hilbert space ${\cal H}_0(\Sigma,\cal P)$. This states
coincide with those defined in \cite{Freidel2} on a fix
triangulation.

\begin{figure}
\psfrag{j}{$j$}
\psfrag{gamma}{$\gamma$}
\psfrag{i3}{$i_3$}
\psfrag{ell1}{$\ell_1$}
\psfrag{ell2}{$\ell_2$}
\psfrag{p1}{$v_{\ell_1}$}
\psfrag{p2}{$v_{\ell_2}$}
\psfrag{jd}{$j$}
\centering
\includegraphics[scale=0.5]{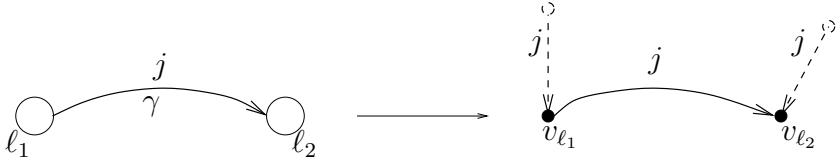}
\caption{\small The two particles spin-network state. Note that the intertwiners associated to the two particles $v_{\ell_1}$ and $v_{\ell_2}$ are different.}
\label{twoparticlestate}
\end{figure}

As an example of generalized spin-network state involving particles degrees of freedom, let us write the two-particles state $\Psi_{\gamma,j}[A;\Lambda_1,\Lambda_2]$ defined by (figure \ref{twoparticlestate}):
\begin{eqnarray}\label{formulatwoparticlestate}
\Psi_{\gamma,j}[A;\Lambda_1,\Lambda_2] \; \equiv \;\; <\stackrel{j}{e}{}\!^{s_1} \vert \stackrel{j}{\pi}{}\!(\Lambda_1^{-1} h_\gamma[A] \Lambda_2) \vert \stackrel{j}{e}{}\!_{s_2}> \;.
\end{eqnarray}
In this formula, $\gamma$ is an oriented path linking the particles $\wp_1$ and $\wp_2$. Physically, this kinematical state represents a system of two self-gravitating particles with the same total angular momentum $j$ and therefore it is identically null if the sum of the spins $s_1+s_2$ is not an integer.

	   \subsubsection*{3.1.3. Spectrum of distance operators}
At this stage, one can already ask the question of the quantization of distance operators (\ref{distanceoperator}) and their action on the generalized kinematical Hilbert space ${\cal H}_0(\Sigma,\cal P)$. In fact, definitions of geometric operators have already been proposed in the framework of pure loop quantum gravity: these operators are self-adjoint operators acting on the kinematical Hilbert space and their spectrum have been computed \cite{Ashtekar, Freidel, Rovelli2}. It is claimed that the nature of these spectra gives the answer of the fundamental question of the space-time geometrical structure at the Planck scale. But this physical interpretation is subject to polemics mainly because these geometric operators are not observables and therefore are not trivially related to a physical process. 

This section aims at clarifying this point, in the three-dimensional framework,
by constructing classical distance observables labeled by a path $\gamma$
linking two particles (\ref{distanceoperator}); we will identify the path and
the associated distance observable in the sequel. Given two such observables,
their Dirac bracket is non-trivial only if their intersection is not empty: if
they share one particle, then the Dirac bracket is quadratic and involves a
solution of CDYB equation \cite{Buffenoir}; otherwise the Dirac bracket is
computed using the Chern-Simons symplectic structure.

\begin{figure}
\psfrag{ell1}{$\ell_1$}
\psfrag{ell2}{$\ell_2$}
\centering
\includegraphics[scale=0.5]{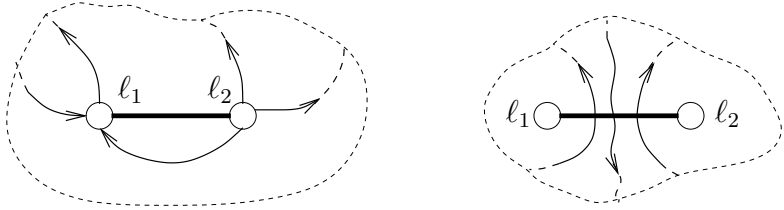}
\caption{\small Distance observables and generalized spin-network states. The distance operator between $\wp_1$ and $\wp_2$ is represented by a thick line; holonomies defining generalized spin-network states are represented by thin lines.}
\label{actionofdistanceoperators}
\end{figure}

In this article, we are exploring a quantization scheme based on loop quantum
gravity techniques. However, the quantization is incomplete in the sense that
we do not quantize geometrical degrees of freedom living on the
boundaries of particles. Therefore we cannot compute the action of a distance operator
between two particles on a kinematical state involving one of these particles
(left picture \ref{actionofdistanceoperators}); this case will be studied in
\cite{Noui2}. It is nevertheless possible to make the computation in any other
cases (right picture \ref{actionofdistanceoperators}) which corresponds to
what it is done in pure loop quantum gravity.

\begin{figure}
\psfrag{gamma}{$\gamma$}
\psfrag{i}{$i$}
\psfrag{j}{$j$}
\psfrag{k}{$k$}
\psfrag{x1}{$\!\!\!x_1$}
\psfrag{x2}{$\!\!\!x_2$}
\centering
\includegraphics[scale=0.5]{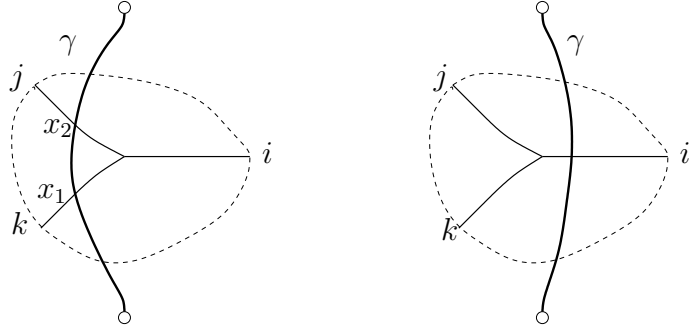}
\caption{\small Action of the distance operator on a spin-network state. The action on the left is weakly equal to the action on the right. By weakly, we mean equal on the constraints surface, i.e. one can set any holonomy (with free ends) of the connection to be trivial on any contractible open set of $\Sigma$.}
\label{Distanceoperatoronspinnetworks}
\end{figure}

Thus, let us consider a classical (square) distance operator
(\ref{distanceoperator}) characterized by a set of any two particles $\{ \wp_i
,\wp_j\}$ and an oriented path $\gamma$ linking them. One can see that there
is no ambiguity to quantize this observable and one obtains a quantum operator
$\hat{\cal{D}}{}^2_\gamma(i,j)$ acting on ${\cal H}_0(\Sigma,\cal P)$. Let us
consider a gauge invariant generalized spin-network state $\Psi_S \in {\cal
H}_0(\Sigma,\cal P)$ associated to a graph $\Gamma$ such that there is one and
only one edge $\zeta \in {\cal E}_\Gamma$ which intersects the curve $\gamma$
on only one point $x \not \in \ell_1,\ell_2$. It is clear that the action of
$\hat{\cal{D}}{}^2_\gamma(i,j)$ on $\Psi$ reduces to the following
action\footnote{Note that if the operator $\hat{\cal D}^2_\gamma(i,j)$ is
clearly a Dirac observable, it is not the case for the operator on the
r.h.s. of the equation (\ref{actionofthedistanceoperator}). Therefore, this
equation is not covariant and one could ask the question of its meaning. In
fact, one has to understand this equation as a gauge fixing of the (square)
distance operator and the r.h.s. of equation
(\ref{actionofthedistanceoperator}) corresponds to the expression of
$\hat{\cal D}^2_\gamma(i,j)$ in a particular gauge. One can make such a gauge
fixing in two different equivalent ways: one can make use of the translational
symmetry to fix to zero the positions of the particles $q_i$ and $q_j$ or one
can make use of the rotational ($SU(2)$)-symmetry to fix the holonomy of the
connection along $\gamma$ to a value such that any gauge transformation
(translations in that case) is trivial. It is important to notice that, due to
the relative configuration between the state $\Psi$ and the distance operator,
these gauge fixing do not affect the state itself. It would have been
different if the state had involved the particle degrees of freedom in the operator. Finally, we
see that the result of the action (\ref{actionofthedistanceoperator}) is gauge
invariant, thus well-defined and does not depend on the choice of the gauge
fixing.}:
\begin{eqnarray}\label{actionofthedistanceoperator}
\hat{\cal{D}}{}^2_\gamma(i,j) \; \rhd \; \Psi \; = \; \hat{q}_\gamma[e,A]^{\dagger} \; \hat{q}_\gamma[e,A] \; \rhd \; \Psi \;.
\end{eqnarray}
Let us recall that ${q}_\gamma[e,A]$ is the translational part of the Chern-Simons holonomy along $\gamma$ and its quantization is defined without any ambiguity by:
\begin{eqnarray}
\hat{q}_\gamma[e,A]^k \; = \; \int_\gamma \!h_{\gamma<x}[\hat{A}]^k{}_l \; \hat{e}_{a}^l(x)dx^{a} \;\; \text{and} \;\; \hat{q}_\gamma[e,A]^\dagger_k \; = \; \int_\gamma  \! h_{\gamma<x}^{-1}[\hat{A}]^l{}_k \; \hat{e}_{a l}(x) dx^{a} \;.
\end{eqnarray}
Therefore, the result of the action (\ref{actionofthedistanceoperator}) is an immediate consequence of the formula (\ref{derivationofholonomy}) and reads:
\begin{eqnarray}
\hat{\cal{D}}{}^2_\gamma(i,j) \; \rhd \; \Psi_S \; = \; s(x,\zeta,\gamma)^2 \; i_\zeta(i_\zeta + 1) \; \Psi \;.
\end{eqnarray}
This result is identical to the usual one obtained in pure loop quantum gravity even if our expression of the distance operator is a priori completely different than the usual one. Nevertheless, contrary to the usual (non-physical) length operator used in loop quantum gravity, this operator is well defined and one does not need to consider a regularization to compute its action. As a result, an important difference occurs when one considers the distance operator (\ref{actionofthedistanceoperator}) acting on states which intersect the path $\gamma$ several times (see left picture of Figure \ref{Distanceoperatoronspinnetworks}). 
In particular, one does not obtain anymore that the spin-network state is an
 eigenstate of the distance operator whose eigenvalue is a sum of
 contributions of different edges intersected. The action seems a priori more
 complicated and not necessary diagonal. However, as the distance operator is
 a Dirac observable, it should be diffeomorphism invariant and therefore
 should depend only on the homotopy class of the path $\gamma$. Thus, one
 could expect to find configurations where the computation of its action on a
 state is simpler and might be diagonal (see Figure
 \ref{Distanceoperatoronspinnetworks}).  
To see this  property, let us consider the example of the Figure
 \ref{Distanceoperatoronspinnetworks}. It is immediate to show that the
 equality between the two actions holds when one assumes that the holonomy of
 the connection along the part of the path $\gamma$ between $x_1$ and $x_2$ is
 trivial, which is indeed the case on the constraints surface. Therefore, we
 have to make a gauge fixing which is state dependent: this point has to be
 related to the fact that the relation between gauge transformations and
 diffeomorphisms involves field-dependent gauge parameters. As a result, the
 action of the distance operator is invariant under local diffeomorphisms
 on-shell. Nevertheless, one has to be careful with global diffeomorphisms
 which does not generically leave the action of the distance operator
 invariant. This point is explained in the companion paper
 \cite{Noui}. Finally, one can really interpret colored edges of any
 spin-network intersecting $\gamma$ as fluxes of geometry and the invariance
 of the action of the distance operator under diffeomorphisms can be viewed as
 the conservation of the total geometry flux toward the path $\gamma$.

Before ending this section, it is important to underline that we gave the
spectrum of distance observables acting on particular states. Given the
distance operator between two particles, the contribution to its spectrum of
generalized spin-network states involving the degrees of freedom of these
particles could be completely different. For instance, it seems clear that the
action of ${\cal D}^2_\gamma(1,2)$ on the two-particle state
(\ref{formulatwoparticlestate}) will not be diagonal because of Heisenberg
uncertainties. Indeed, such an action can be physically interpreted as the
result of a measure of the distance between the two particles whereas the
momenta of each particle are completely determined. The article \cite{Noui2}
aims to clarify these aspects.

         \subsection*{3.2. Spin-Foam amplitude}
This section is devoted to implement the remaining first class constraints
$\Phi_C$ (\ref{separationofconstraints}) in order to define the Physical
Hilbert space ${\cal H}_{phys}(\Sigma,{\cal P})$ of the coupled system. The
idea consists of generalizing the construction presented in \cite{Noui} by
introducing a projection operator on physical states in the presence of
particles. Then, we show that the physical scalar product admits a spin-foam
state sum representation which defines a Ponzano-Regge model in the presence
of massive spinning particles. Finally, we discuss the possibility to extend
this construction to have a description of a background independent quantum
field theory.

	    \subsubsection*{3.2.1. Physical Hilbert space: projector and physical scalar product}
Given any function $v \in C^{\infty}(\Sigma,\mathbb R^3)$, the functional $\Phi_C(v)$ imposes constraints on the Dirac classical phase space of the coupled system. The constraints are first class if and only if the function $v$ is constant on each boundaries $\ell_i$; otherwise they are second class \cite{Buffenoir}. Thus, it is clear that the second class part of $\Phi_C(v)$ imposes that the component $A_{\varphi_i}$ of the spin-connection along each circle $\ell_i$ is constant and this condition has been already implicitly considered at the kinematical level. On the other hand, the first class part of the functional $\Phi_C(v)$ imposes the condition that the spin-connection is flat everywhere on $\Sigma$ and fixes its components $A_{\varphi_i}$ on each circle $\ell_i$ to the value:
\begin{eqnarray}\label{boundaryconstraint}
\forall \; x \; \in \; \ell_i \;\;, \;\;\; A_{\varphi_i}(x) \; \equiv \; A_{\varphi_i}(\ell_i) \; = \; -\frac{1}{2\pi} m_i \; \Lambda_i \; J_0 \; \Lambda_i^{-1} \;.
\end{eqnarray}

At the quantum level, imposing these constraints on the generalized kinematical Hilbert space should schematically identify spin-network states associated to graphs which are related by a diffeomorphism (connected to the identity). This is exactly what happens in the pure gravitational case \cite{Noui}. Indeed, we construct an orthonormal basis ${\cal B}=(\Omega_J)_J$ of the physical Hilbert space ${\cal H}_{phys}(\Sigma)$ labeled by the so-called irreducible graph colored by a family $J=(j_1,\cdots,j_{6g-6})$ of irreducible unitary representations of $SU(2)$ ($g>1$ being the genus\footnote{The case of the sphere and the torus are treated separately but do not introduce particular difficulties.} of the surface $\Sigma$). Any kinematical spin-network state $\Psi_S$ labeled by a spin-network $S=(\Gamma,(j_\gamma)_{\gamma},(\iota_v)_v)$ can be projected to the physical Hilbert space and its components in the basis $\cal B$ are given by:
\begin{eqnarray}\label{physicalscalarproduct}
c_{IS} \; \equiv \; <\Psi_S, \Omega_I>_{phys} \; = \; <P\Psi_S , \Omega_I> \;,
\end{eqnarray}
where $<,>_{phys}$ is the physical scalar product defined from the kinematical scalar product $<,>$ and the projection operator $P$ on physical states acting on ${\cal H}_0(\Sigma)$ \cite{Noui}. To be more precise, the operator $P$ is not well-defined in ${\cal H}_0(\Sigma)$ in the sense that it sends any kinematical state to an element of $\text{Cyl}^{\star}(\Sigma)$, the algebraic dual of $\text{Cyl}(\Sigma)$. Therefore, $P$ is not strictly speaking a projector and the physical scalar product (\ref{physicalscalarproduct}) is in fact a bilinear form between $\text{Cyl}(\Sigma)$ and $\text{Cyl}^{\star}(\Sigma)$. Here, we have implicitly identified by duality the basis of $\text{Cyl}(\Sigma)$ with its dual basis.

From the definition (\ref{physicalscalarproduct}), we introduce an equivalence relation between kinematical spin-network states as follows:
\begin{eqnarray}
\Psi_S \; \sim \; \Psi_{S'} \;\;\; \text{if and only if} \;\;\; \forall \; J \;,\;\; <\Psi_S - \Psi_{S'},\Omega_J>_{phys} \; = \; 0 \;.
\end{eqnarray}
One says that the two states $\Psi_S$ and $\Psi_{S'}$ are physically equivalent. Given two spin-networks $S_1$ and $S_2$ such their associated graphs $\Gamma_1$ and $\Gamma_2$ are related by a diffeomorphism, one can show that $\Psi_{S_1} \sim \Psi_{S_2}$ \cite{Noui}. This property clarifies the above claiming about the fact that the constraints identify diffeomorphically equivalent spin-networks (as one could naturally expect). 

\medskip

\begin{figure}
\psfrag{ell1}{$\ell_1$}
\psfrag{ell2}{$\ell_2$}
\psfrag{phi11}{$\varphi_1^{(1)}$}
\psfrag{phi12}{$\varphi_1^{(2)}$}
\psfrag{phi13}{$\varphi_1^{(3)}$}
\psfrag{phi14}{$\varphi_1^{(4)}$}
\psfrag{phi21}{$\varphi_2^{(1)}$}
\psfrag{phi22}{$\varphi_2^{(2)}$}
\psfrag{phi23}{$\varphi_2^{(3)}$}
\psfrag{phi24}{$\varphi_2^{(4)}$}
\centering
\includegraphics[scale=0.7]{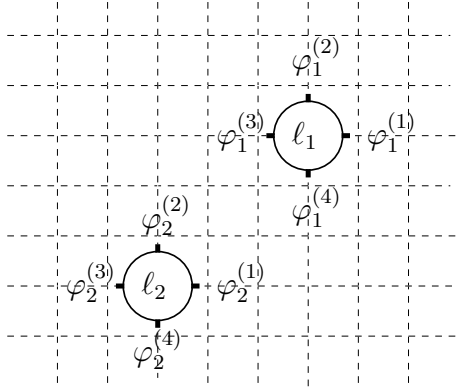}
\caption{\small Cellular decomposition of the surface $\Sigma$ with boundaries. For convenience reasons, we have drawn a regular square lattice. It induces natural simplicial decompositions of the boundaries associated to particles.}
\label{cellulardecomposition}
\end{figure}

This section aims to generalize the previous construction to the coupled system. In order to define a projection operator on physical states, it will be convenient to consider, as in the pure gravitational case, a cellular decomposition of the surface with boundary $\Sigma$ such that each circle is contractible and does not contain any particle. The lattice intersects each boundary $\ell_i$ in $\nu_i$ different points denoted $(\varphi_i^{(1)},\cdots,\varphi_i^{(\nu_i)})$ (figure \ref{cellulardecomposition}). As a result, it induces a decomposition of each boundary $\ell_i$ in $\nu_i$ parts and we will denote by $e_i^{(n)}$ the part going from $\varphi_i^{(n)}$ to $\varphi_i^{(n+1)}$. In the sequel, the set of edges and faces of the cellular decomposition will be respectively denoted by ${\cal S}^1(\Sigma)$ and ${\cal S}^2(\Sigma)$.
To each element of $e \in {\cal S}^1(\Sigma)$, we assign the $SU(2)$-group element $h_e[A]$ and we will denote by $U_f[A]=\prod_{e \in f} h_e[A] \in SU(2)$ the holonomy of the spin-connection around the face $f$. Using these notations, we define the distributions $P_{\ell_i}[A,\Lambda_i] = \prod_{n=1}^{\nu_i}  P_{\ell_i}^{(n)}$ where: 
\begin{eqnarray}
 P_{\ell_i}^{(n)}[A,\Lambda_i] \; \equiv \; \delta(h_{e_i^{(n)}}[A] \; \Lambda_i \exp(m_i \frac{\varphi_{i}^{(n+1)}-\varphi_{i}^{(n)}}{2\pi}J_0)\Lambda_i^{-1}) \;\;.
\end{eqnarray}
It is clear that the distributions ${P}_{\ell_i}$ are a discretize version of the constraints (\ref{boundaryconstraint}). As a consequence, it seems natural to propose the following lattice version for the whole first class constraint $\Phi_C$:
\begin{eqnarray}\label{projectorforcoupledsystem}
P[A,(\Lambda_i)_i] \; \equiv \; \prod_{f \in {\cal S}^2(\Sigma)} \delta(U_f[A]) \; \prod_{i=1}^N P_{\ell_i}[A,\Lambda_i] \;.
\end{eqnarray}
There is no ambiguity in quantizing this operator for it involves only commuting variables. However, the obtained quantum operator $\hat{P}$ is clearly not well-defined in the generalized kinematical Hilbert space ${\cal H}_0(\Sigma,\cal P)$ for the same reasons as in the pure gravitational case \cite{Noui}. Therefore, $\hat{P}$ is not, strictly speaking, a projector but defines an application from ${\cal H}_0(\Sigma,\cal P)$ to the space $\text{Cyl}^{\star}(\Sigma,\cal P)$, the algebraic dual of the space of generalized cylindrical functions (\ref{spaceofgeneralizedcylindricalfunctions}). Nevertheless, we will identify in the sequel any element $\Psi_S$ of a generalized spin-network basis in $\text{Cyl}(\Sigma,\cal P)$ to its dual generalized spin-network in $\text{Cyl}^\star(\Sigma,\cal P)$. Following the usual construction, we define the generalized physical scalar product between any two generalized spin-network states $\Psi_S$ and $\Psi_{S'}$ by the expression:
\begin{eqnarray}\label{generalizedphysicalscalarproduct}
<\Psi_S,\Psi_{S'}>_{phys} \; \equiv \; <P\Psi_S,\Psi_{S'}> \;.
\end{eqnarray}
Let us recall that $<,>$ is the scalar product on the Hilbert space ${\cal H}_0(\Sigma,{\cal P})$. The expression (\ref{generalizedphysicalscalarproduct}) is defined for any cellular decomposition and therefore we can compute the scalar product between any two spin-network states based on any (continuous) graph immersed in the surface $\Sigma$.

Two generalized spin-network states $\Psi_S$ and $\Psi_{S'}$ are called physically equivalent ($\Psi_S \sim \Psi_{S'}$) if and only if they satisfy the following property:
\begin{eqnarray}
\forall \;  F\; \in \; {\cal H}_0(\Sigma,{\cal P}) \;\;,\;\;\;\; <\Psi_S - \Psi_{S'},F>_{phys} \; = \; 0 \;.
\end{eqnarray}
To have a geometrical interpretation of this equivalence relation, let us consider two generalized spin-networks $S_1$ and $S_2$ whose associated graphs $\Gamma_1$ and $\Gamma_2$ are related by a continuous diffeomorphism $\phi$ (connected to the identity). If the graphs do not touch the boundaries, then the spin-network states $\Psi_{S_1}$ and $\Psi_{S_2}$ are physically equivalent as in the pure gravitational case. However, we cannot generically conclude to the physical equivalence between the spin-network states when the graphs do touch the boundaries as we show in the following example.

\begin{figure}
\psfrag{ell}{$\ell_i$}
\psfrag{phi1}{$\varphi_1$}
\psfrag{phi2}{$\varphi_2$}
\psfrag{theta1}{$\vartheta_1$}
\psfrag{theta2}{$\vartheta_2$}
\psfrag{phi}{$\phi$}
\psfrag{j}{$j$}
\psfrag{k}{$k$}
\psfrag{0}{$0$}
\centering
\includegraphics[scale=0.7]{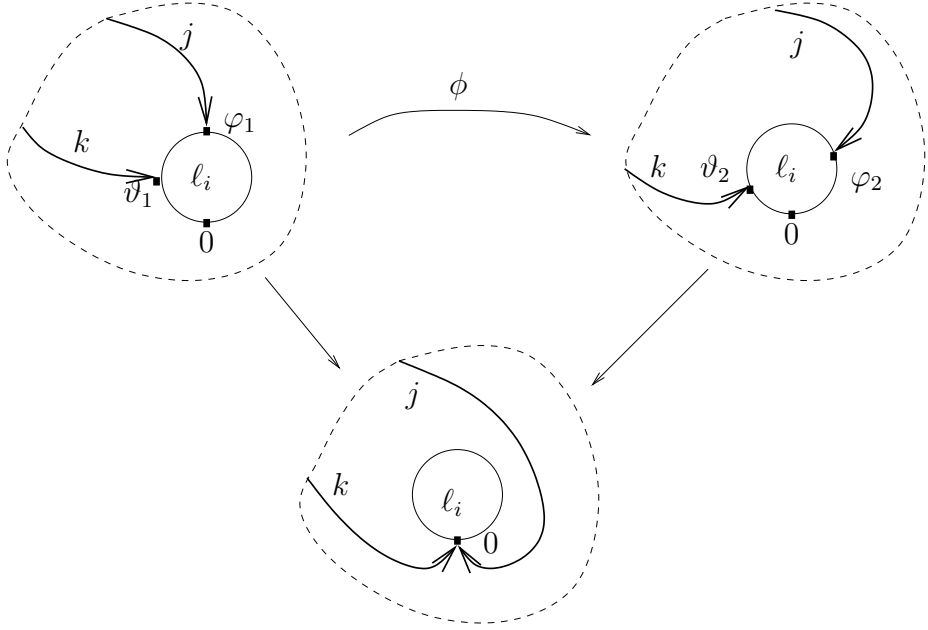}
\caption{\small Example of two spin-networks whose graphs are related by a continuous diffeomorphism $\phi$. The points $\varphi_1$ and $\vartheta_1$ are respectively sent to the points $\varphi_2$ and $\vartheta_2$ by this diffeomorphism. The associated spin-network states $\Psi_{S_1}$ and $\Psi_{S_2}$ are physically equivalent to linear combinations of spin-network states associated to the graph on the bottom.}
\label{boundarydiffeo}
\end{figure}

Let us consider a generalized spin-network $S_1$ whose graph touches (only) one of the circles $\ell_i$ in two different points $\varphi_1$ and $\vartheta_1$; the graph associated to the spin-network $S_2$ intersects the same circle at the points $\varphi_2$ and $\vartheta_2$ (figure \ref{boundarydiffeo}).
The spin-network states $\Psi_{S_\alpha}$ ($\alpha=1,2$) associated to these graphs can be written as follows:
\begin{eqnarray}
\Psi_{S_{\alpha}}[A,\Lambda_i] \; = \; R^{(S_\alpha)}_{ab}\; \stackrel{j}{\pi}{}\!(h_{\gamma_{\alpha}}[A])^a_{a'} \; \stackrel{k}{\pi}{}\!(h_{\zeta_{\alpha}}[A])^b_{b'} \;  \stackrel{j_{\ell_i}}{\pi}{}\!(\Lambda_i)^{c'}_{s_i} \;\; \iota_{\ell_i}(j,k,j_{\ell_i})^{a'b'}_{c'} \;.
\end{eqnarray}
In this expression, we have used the notation $ R^{(S_\alpha)}_{ab}$ for the functions associated to the rest of the spin-networks drawn in (figure \ref{boundarydiffeo}). Let us recall that $(j_{\ell_i},\iota_{\ell_i})$ are the color associated to the boundary $\ell_i$; in that particular case, there is a unique choice (up to a normalization) for the intertwiner $\iota_{\ell_i}\equiv \iota_{\ell_i}(j,k,j_{\ell_i}): \stackrel{j}{V} \otimes \stackrel{k}{V} \otimes \stackrel{j_{\ell_i}}{V} \rightarrow \mathbb C$.

Actually, the spin-network state $\Psi_{S_1}$ is physically equivalent to a linear combination of spin-network states $\Psi_{S'_2}$ whose spin-networks $S'_2$ differ from $S_2$ only by the color associated to the boundary $\ell_i$ as it is shown in the following calculation:
\begin{eqnarray}
\Psi_{S_1}[A,\Lambda_i] & = & R^{(S_1)}_{ab}\; \stackrel{j}{\pi}{}\!(h_{\gamma_1}[A] \Lambda_i)^a_{a'} \; \stackrel{k}{\pi}{}\!(h_{\zeta_1}[A]\Lambda_i)^b_{b'} \; \iota_{\ell_i}(j,k,j_{\ell_i})^{a'b'}_{s_i} \nonumber \\
& \sim &   R^{(S_1)}_{ab}\; \stackrel{j}{\pi}{}\!(h_{\gamma_2}[A] h_{\gamma_{2\rightarrow 1}}[A]\Lambda_i)^a_{a'} \; \stackrel{k}{\pi}{}\!(h_{\zeta_2}[A]h_{\zeta_{2\rightarrow 1}}[A] \Lambda_i)^b_{b'} \; \iota_{\ell_i}(j,k,j_{\ell_i})^{a'b'}_{s_i} \nonumber \\
& \sim &  R^{(S_1)}_{ab}\; \stackrel{j}{\pi}{}\!(h_{\gamma_2}[A] \Lambda_i e^{m_i\frac{\Delta \vartheta}{2\pi}J_0})^a_{a'} \; \stackrel{k}{\pi}{}\!(h_{\zeta_2}[A] \Lambda_i e^{m_i\frac{\Delta \varphi}{2\pi}J_0})^b_{b'} \; \iota_{\ell_i}(j,k,j_{\ell_i})^{a'b'}_{s_i} \nonumber \\
& \sim & R^{(S_1)}_{ab}\; \stackrel{j}{\pi}{}\!(h_{\gamma_2}[A] \Lambda_i)^a_{a'} \; \stackrel{k}{\pi}{}\!(h_{\zeta_2}[A] \Lambda_i)^b_{b'} \; e^{i m_i \frac{a'\Delta \vartheta + b' \Delta \varphi}{2\pi}}\;\iota_{\ell_i}(j,k,j_{\ell_i})^{a'b'}_{s_i} \nonumber \\
& \sim & \sum_{S'_2} c_{S_1 S'_2} \; \Psi_{S'_2}[A,\Lambda_i] \; . \nonumber
\end{eqnarray} 
In these lines of arguments, we have introduced the notations $h_{\gamma_{2\rightarrow 1}}[A]$ (resp. $h_{\zeta_{2\rightarrow 1}}[A]$) for the holonomy of the connection along $\ell_i$ between $\vartheta_1$ and $\vartheta_2$ (resp. $\varphi_1$ and $\varphi_2$); $\Delta \vartheta = \vartheta_2 - \vartheta_1$ and $\Delta \varphi = \varphi_2 - \varphi_1$. This calculation makes use of the facts that $\iota(j,k,j_{\ell_i})$ is an intertwiner, $h_{\gamma_{2\rightarrow 1}}[A]$ and $h_{\zeta_{2\rightarrow 1}}[A]$  are physically equivalent to $e^{m_i\frac{\Delta \vartheta}{2\pi}J_0}$ and $e^{m_i\frac{\Delta \varphi}{2\pi}J_0}$, and $R^{(S_1)}_{ab}$ is physically equivalent to $R^{(S_2)}_{ab}$. The last line results from the orthogonality relation $\sum_{j'_{\ell_i}} \iota(j,k,j'_{\ell_i}) \iota(j,k,j'_{\ell_i})=\stackrel{j}{\pi}\!\!(1)\otimes \stackrel{k}{\pi}\!\!(1) $ and $c_{S_1S'_2}$ is a coefficient easy to compute. 

Of course, one can generalize the previous result to any spin-network state. In particular, any generalized spin-network can be physically extended as a sum of generalized spin-networks which touch the boundaries at their origin (figure \ref{boundarydiffeo}). For that reason, we will exclusively consider in the sequel latter types of generalized spin-networks.

         \subsubsection*{3.2.2. Ponzano-Regge model in presence of particles}
This section aims to picture the physical scalar product between any two spin network states $\Psi_S$ and $\Psi_{S'}$ in terms of spin-foam models. As in the pure gravitational case \cite{Noui}, we will make use of this picture to construct, in a second step, an explicit basis of the physical Hilbert space. For convenience reasons, we will not give a general expression for the physical scalar product but we will illustrate the construction through typical examples which provide in fact a recipe valid for any situation. 

Before going to the coupled system, let us briefly review the main steps of the construction in the pure gravitational case \cite{Noui}. Given two spin network states $\Psi_S$ and $\Psi_{S'}$ respectively defined on the graphs $\Gamma_{S}$ and $\Gamma_{S'}$, we define the graph $\Gamma_{SS'}$ to be the minimal graph that contains $\Gamma_{S}$ and $\Gamma_{S'}$. To this graph is associated the notion of set of irreducible loops $\alpha^{SS'}$ defined as the set of oriented boundaries of the corresponding graph. With these definitions, the physical scalar product reads:
\begin{eqnarray}\label{puregravitationalscalarproduct}
<\Psi_S , \Psi_{S'}>_{phys} \; = \; <\Psi_{S}, \prod_{\gamma \in \alpha^{SS'}} \delta({U}_\gamma[A]) \; \Psi_{S'}> \;.
\end{eqnarray}
The operator ${U}_\gamma[A]$ denotes the holonomy of the connection around the irreducible loop $\gamma$ and acts on the states by multiplication. Then, we finish the calculation using the Ashtekar-Lewandowski measure. 

For the coupled system, the picture is similar but we use the expression (\ref{projectorforcoupledsystem}) for the projector. Thus, the expression of the physical scalar product between two generalized spin network states differs from the pure gravitational case only when the states involve particles degrees of freedom or when there exists an irreducible loop surrounding one or several particles. The following aims to show the construction of the physical scalar product in the latter cases.

\begin{figure}
\psfrag{ell}{$\ell_i$}
\psfrag{iota}{$\iota_{\ell_i}$}
\psfrag{ji}{$j_{\ell_i}$}
\psfrag{a}{$(a)$}
\psfrag{b}{$(b)$}
\psfrag{c}{$(c)$}
\psfrag{j}{$j$}
\psfrag{k}{$k$}
\psfrag{l}{$l$}
\centering
\includegraphics[scale=0.7]{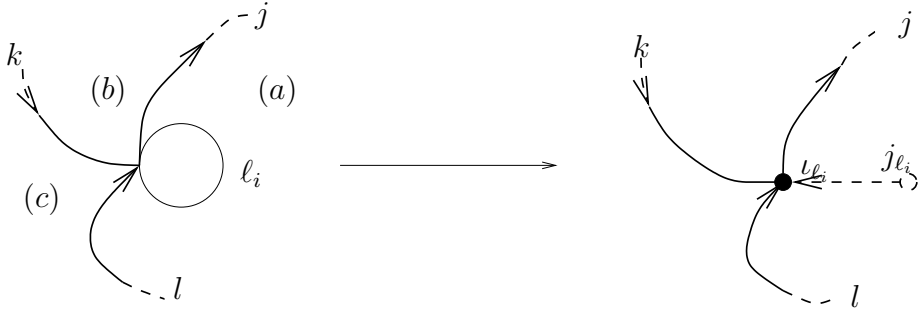}
\caption{\small Example of a generalized spin-network involving particles degrees of freedom. The vicinity of the particle is divided into three regions $(a)$, $(b)$ and $(c)$ by the graph associated to this spin-network (on the right). The region $(a)$ is called the particle area and the spin-network state (on the left) is defined such that its ``virtual'' particle is located in the particle area.}
\label{regularizationofspinnetworks}
\end{figure}

For that purpose, let us consider a state involving the degrees of freedom of one particle $\wp_i$ (figure \ref{regularizationofspinnetworks}). The spin-network $S$ associated to this state is such that its graph $\Gamma_S$ touches the boundary $\ell_i$ only at the origin and therefore separates the vicinity of the particle into different areas. Only one of these areas contains the particle and we will refer to it as the particle area. As a result, we define any spin-network state in such a way that the ``virtual'' particle is located in the particle area.
Given two such spin-networks $S$ and $S'$, we can still define the set of irreducible loops $\alpha^{SS'}$ in the same way as in the pure gravitational case. Nevertheless, we have to distinguish the subset $\alpha_1^{SS'}$ of irreducible loops surrounding particle areas from the subset $\alpha_0^{SS'}$ of irreducible loops associated to empty areas. In that regularization, the projector operator $P$ reads as follows:
\begin{eqnarray}\label{pito}
P \; = \; \prod_{\gamma \in \alpha^{SS'}_0} \delta(U_\gamma[A]) \prod_{\zeta_i \in \alpha^{SS'}_1} \delta(U_{\zeta_i}[A]\Lambda_i e^{m_iJ_0} \Lambda_i^{-1}) \;.
\end{eqnarray}
This expression of $P$ is a trivial consequence of the definition (\ref{projectorforcoupledsystem}) and $\zeta_i$ denotes the irreducible loop surrounding the particle $\wp_i$. This operator acts naturally by multiplication on any spin-network state. To compute its matrix elements,  one proceeds as follows: first, one makes use of the $SU(2)$ Peter-Weil decomposition formula of the Dirac distribution $\delta(x) = \sum_j \Delta_j \chi_j(x)$ in terms of characters $\chi_j(x)$ of $SU(2)$ finite dimensional representations; then, one uses the generalized kinematical scalar product to finish the computation. The last point is performed using graphical techniques (see appendix of \cite{Noui}).

\begin{figure}
\psfrag{j}{$j$}
\psfrag{iota}{$\iota$}
\psfrag{iotap}{$\iota'$}
\psfrag{k}{$k$}
\psfrag{l}{$l$}
\psfrag{m}{$i$}
\psfrag{p}{$p$}
\psfrag{q}{$q$}
\psfrag{r}{$r$}
\psfrag{kp}{$k'$}
\psfrag{lp}{$l'$}
\psfrag{mp}{$i'$}
\psfrag{jp}{$j'$}
\psfrag{s}{$t$}
\psfrag{left}{$<$}
\psfrag{right}{$>_{phys}$}
\psfrag{mid}{$,$}
\psfrag{=}{$=$}
\psfrag{delta}{$\frac{\delta_{ii'} \delta_{kk'} \delta_{ll'}}{\Delta_i \Delta_k \Delta_l} \sum_t$}
\centering
\includegraphics[scale=0.5]{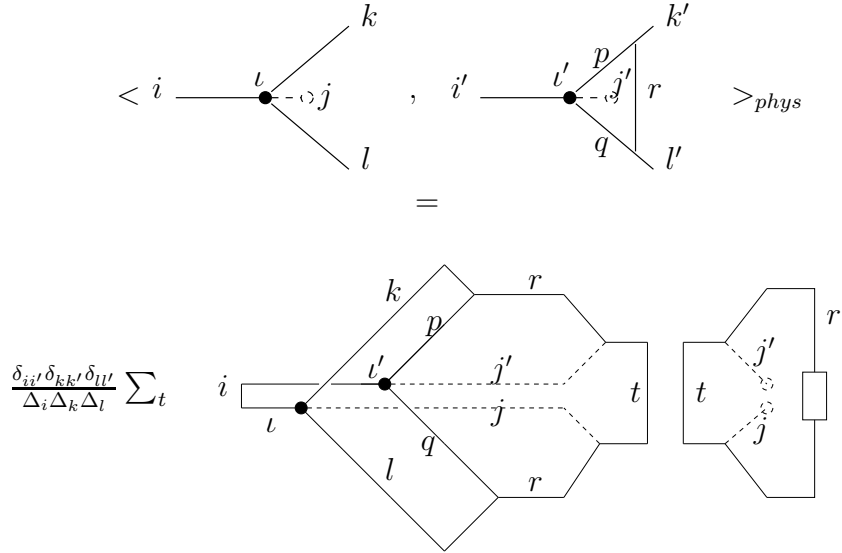}
\caption{\small Physical scalar product between two states. The graphical rules are given in the appendix of \cite{Noui} and the box represents the group element $e^{mJ_0}$ where $m$ is the mass of the particle. The obtained result gives the amplitude of the Ponzano-Regge model associated to massive spinning particles. The presence of the dimensions $\Delta_i$ is just a matter of normalization. Note that the spin $s$ of the particle is contained in the expression of the graph on the right (it labels the magnetic numbers of the open edges).}
\label{twostatesphysicalscalarproduct}
\end{figure}

\begin{figure}
\psfrag{j}{$j$}
\psfrag{iota}{$\iota$}
\psfrag{iotap}{$\iota'$}
\psfrag{k}{$k$}
\psfrag{l}{$l$}
\psfrag{m}{$i$}
\psfrag{p}{$p$}
\psfrag{q}{$q$}
\psfrag{r}{$\;\;r$}
\psfrag{kp}{$k'$}
\psfrag{lp}{$l'$}
\psfrag{mp}{$i'$}
\psfrag{jp}{$j'$}
\psfrag{i}{$i$}
\centering
\includegraphics[scale=0.5]{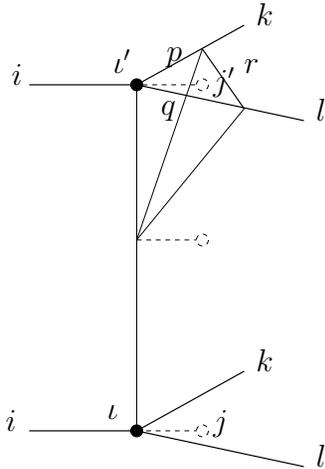}
\caption{\small Spin-foam picture of the transition amplitude between two generalized spin-network states. Note that, in that example, the particle face is the one colored by the representation $r$.}
\label{spinfoampicture}
\end{figure}

To illustrate this construction, let us compute the physical scalar product between the two states $\Psi_S$ and $\Psi_{S'}$ pictured in (figure \ref{twostatesphysicalscalarproduct}). 
The result gives the amplitude of the generalized Ponzano-Regge model in
the presence of massive spinning particles. When one considers the
normalization factors, the transition amplitude ${\cal A}(\Psi_S,\Psi_{S'})$
between the states $\Psi_S$ and $\Psi_{S'}$ is given by the following formula:
\begin{eqnarray}
{\cal A}(\Psi_S,\Psi_{S'}) \; = \; \frac{<P \Psi_S,\Psi_S'>}{\sqrt{<\Psi_S,\Psi_S><\Psi_{S'},\Psi_{S'}>}} \;,
\end{eqnarray}
and can be pictured in terms of spin-foam model as shown in the picture (figure \ref{spinfoampicture}). In that picture, one can see that the particle is associated to an edge but also to a face of the spin-foam. Indeed, the particle evolves along an edge but at the level of the vertex it chooses to follow one of the three emerging edges and therefore selects one of the three emerging faces (dual to edges). The selected face is called the particle face.

In the case where there is no particle ($m=0$ and $s=0$), the amplitude of the
vertex reduces to the usual $(6j)$-symbol and we recover the well-known
Ponzano-Regge model. In the case where the representations $j$ and $j'$
coloring the particle are trivial (particle without total angular momentum),
the spin-network states do not explicitely involve the particle degrees of
freedom but the definition of the projector $P$ does: as a result, the
amplitude of the vertex is still given by the $(6j)$-symbol but the amplitude
of the particle face is modified. To be more precise, the face in the
spin-foam colored by the representation $r$ carries the weight
$\chi_r(e^{mJ_0})$ instead of the dimension $\Delta_r$: this face contains the
information that there is a massive particle. This result is analogous to the
one obtained in the covariant formulation of \cite{Freidel2}. Finally, one can define the model in the general case: the result depends on the choice of the intertwiners $\iota$ and $\iota'$, and the amplitude is given by a function labeled by $11$ representations. 

\medskip

\begin{figure}
\centering
\includegraphics[scale=0.7]{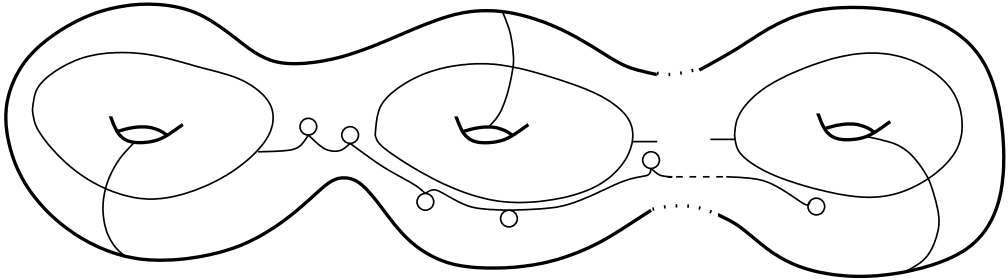}
\caption{\small A basis of the physical Hilbert space for a surface of arbitrary genus $g$ and a number $N$ of particles on it.}
\label{basisofphysicalHilbertspace}
\end{figure}

At this stage, one can make use of the picture of the physical scalar product
in terms of spin foam amplitudes to construct a basis of the physical Hilbert
space when $\Sigma$ is a Riemann surface of arbitrary genus $g$ with an
arbitrary number $N$ of particles. In fact, the construction mimics the pure
gravitational case \cite{Noui} and the result is illustrated in the picture
(figure \ref{basisofphysicalHilbertspace}) for a surface of genus $g>1$ with a
number $N$ of particles: we link all the particles together and we add the
obtained chain to the graph associated to the pure gravitational case. One can
easily see that any element of the basis is labeled by $6g-6+2N$
representations.  Furthermore, one can show that any generalized spin-network
state is physically equivalent to a linear combination of elements of this
basis: in the spin-foam language, this property says that one can ``evolve'',
by the action of the projector $P$, any spin-network state in ${\cal
H}_0(\Sigma, \cal P)$ to a linear combination of ``simplest'' spin network
states based on graphs where the maximal set of irreducible loops have been
annihilated. Note that the choice of the basis is not canonical in the sense
that there exist different parametrizations of the basis: for instance, one
can glue the chain of particles to any edge (associated to pure gravitational
degrees of freedom); or one can cut the chain in two parts and glue them on
different edges... All these parametrizations give equivalent basis of the
physical Hilbert space ${\cal H}_{phys}(\Sigma,\cal P)$.

The case of the torus is similar: the basis is labeled by a graph obtained by adding to the pure gravitational basis graph \cite{Noui} the chain of particles. Any element of the basis is labeled by $2+2N$ representations. The case of the sphere is simpler: the basis is labeled by the chain of particles. When there is only one particle, the basis is trivial; when there are two or three particles, the basis is respectively labeled by $1$ or $3$ representations; when the number of particles is $N>3$, then the basis is labeled by $2N+3$ representations.


\section*{4. Conclusion and perspectives}
In this article, we began by setting out the classical coupling
between three dimensional Euclidean gravity and point particles. We
have briefly presented the canonical analysis of the coupled system
and we have expressed distance operators between any two particles as
Dirac observables. Then, we have adapted loop quantum gravity
techniques to quantize the coupled system: $(i)$ we have generalized
the notion of spin-network to provide the kinematical Hilbert space of
the coupled system with an orthonormal basis; $(ii)$ we have
constructed a basis of the physical Hilbert space and the physical
scalar product can be pictured in terms of spin-foam models. As a
result, we propose a generalization of the Ponzano-Regge model which
includes massive spinning point particles. Furthermore, the
distance observables have been quantized and we have shown that their
action on spin-network states reproduces the well known result in
three dimensional loop quantum gravity using non-physical distance
operators. 

Even if we worked in the framework of Euclidean gravity without
cosmological constant, most of our results can be easily adapted to
the Lorentzian case and can include a cosmological constant. In all
the cases, the classical dynamics of the coupled system is governed by
the same action as the one we used but the gauge group is different
and depends on the signature of the metric and the sign of the
cosmological constant \cite{Witten}. Up to some technical subtleties
\cite{Buffenoir}, one can define distance observables between two
particles in a similar way. The quantization scheme is the same in the
sense that one has to construct the kinematical Hilbert space and then
the physical Hilbert space using spin-foam techniques. However, the
analytic issues are slightly more involved in the Lorentzian case
because of the non-compactness of gauge groups. As for the inclusion
of a cosmological constant in the model, it should turn classical
groups into quantum groups whose deformation parameter $q$ explicitely
depends on the cosmological constant (see \cite{BNR} for example).

One of the main aspects of this article is certainly the quantization
of distance observables and their evaluation on particular generalized
spin-network states. Our distance operator is a well-defined operator on spin-network states and does not need any type of regularization, contrary the usual (non-physical) length operator used in loop quantum gravity. Its action on particular spin-network states have been computed and the result is slightly different than the usual one. All the same, the spectrum is still discrete.
The algebra of distance operators has
not been studied in details but certainly deserves a deeper
understanding. In particular, it would be interested to compute the
action of the distance operator between two particles on the
associated two-particles state (figure \ref{twoparticlestate}): due to
Heisenberg uncertainties, it is clear that this action will not be
diagonal. Usual loop quantum gravity does not consider such situations
and therefore it is not trivial to predict whether they contribute to
a continuous or discrete spectrum. Whatever the result is, it will be
certainly very interesting.

A novel proposal providing a conceptual framework for background
independent quantum field theories has been put forward in
\cite{carloro}. The essential idea is the idealization of a
measurement in a (background independent) quantum mechanical context
by the specification of quantities on the boundary of a local
space-time region. The physical content in the boundary
(e.g. its (space-time) geometry and matter field content) must be specified by a set
(infinite in a theory with local excitations) of Dirac observables
defined on the boundary Hilbert space. Because Dirac observables are
very difficult to construct in four dimensions this proposal is
presented at a formal level. Here we want to point out that a complete
specification of the (physical) geometry of the boundary can be
constructed in terms of the Dirac observables---introduced in section
(2.2.2.)---that define physical distance between point particles.
\begin{figure}[h!!!!!!!!!!!!!!!!!!!!!!!!!!!!!!!!!!!!!!!!!!!!!!!!]
 \centerline{\hspace{0.5cm}\(
\begin{array}{ccc}
\includegraphics[height=6cm]{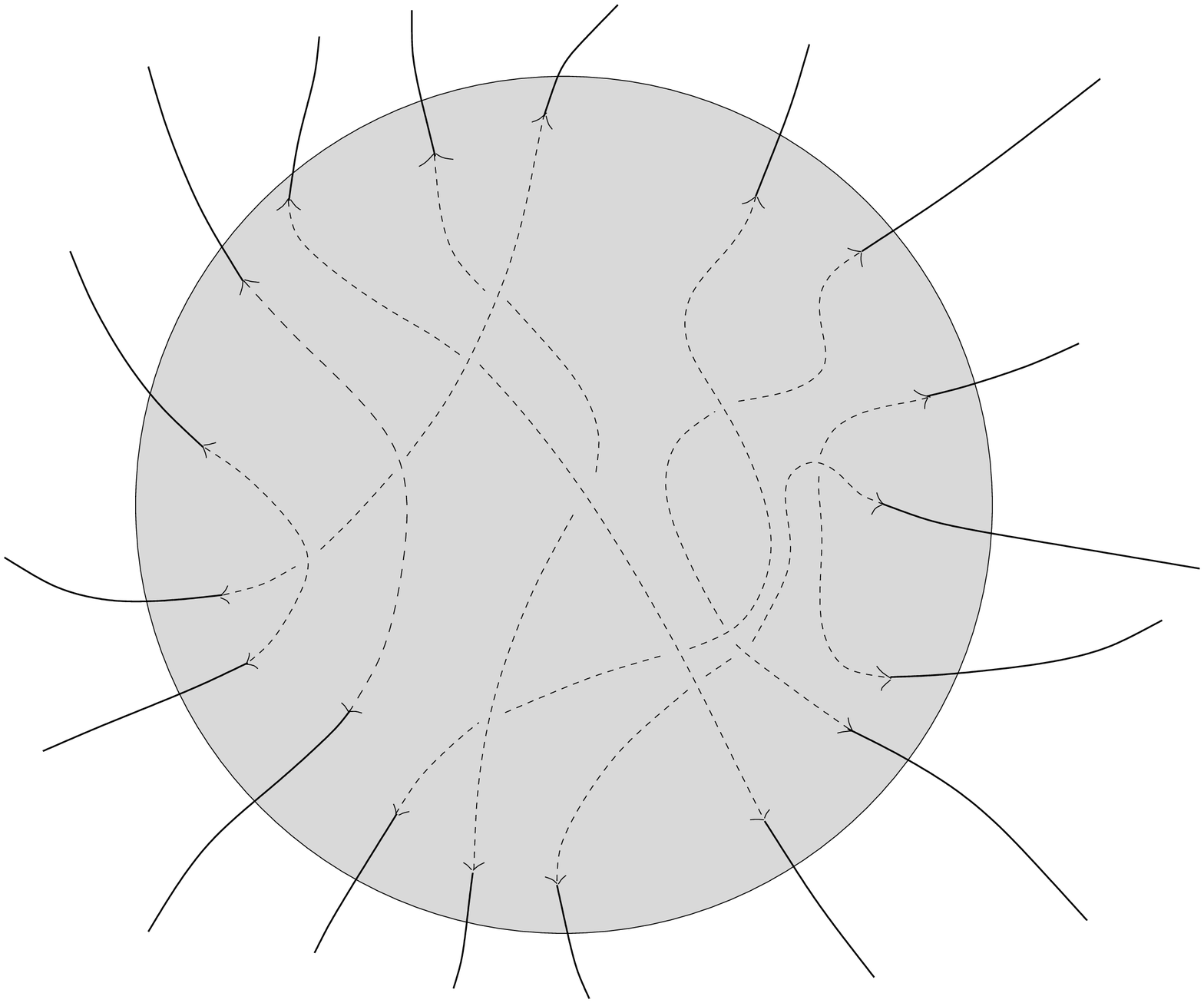} \ \   \ \ \ \ \ \ 
\includegraphics[height=6cm]{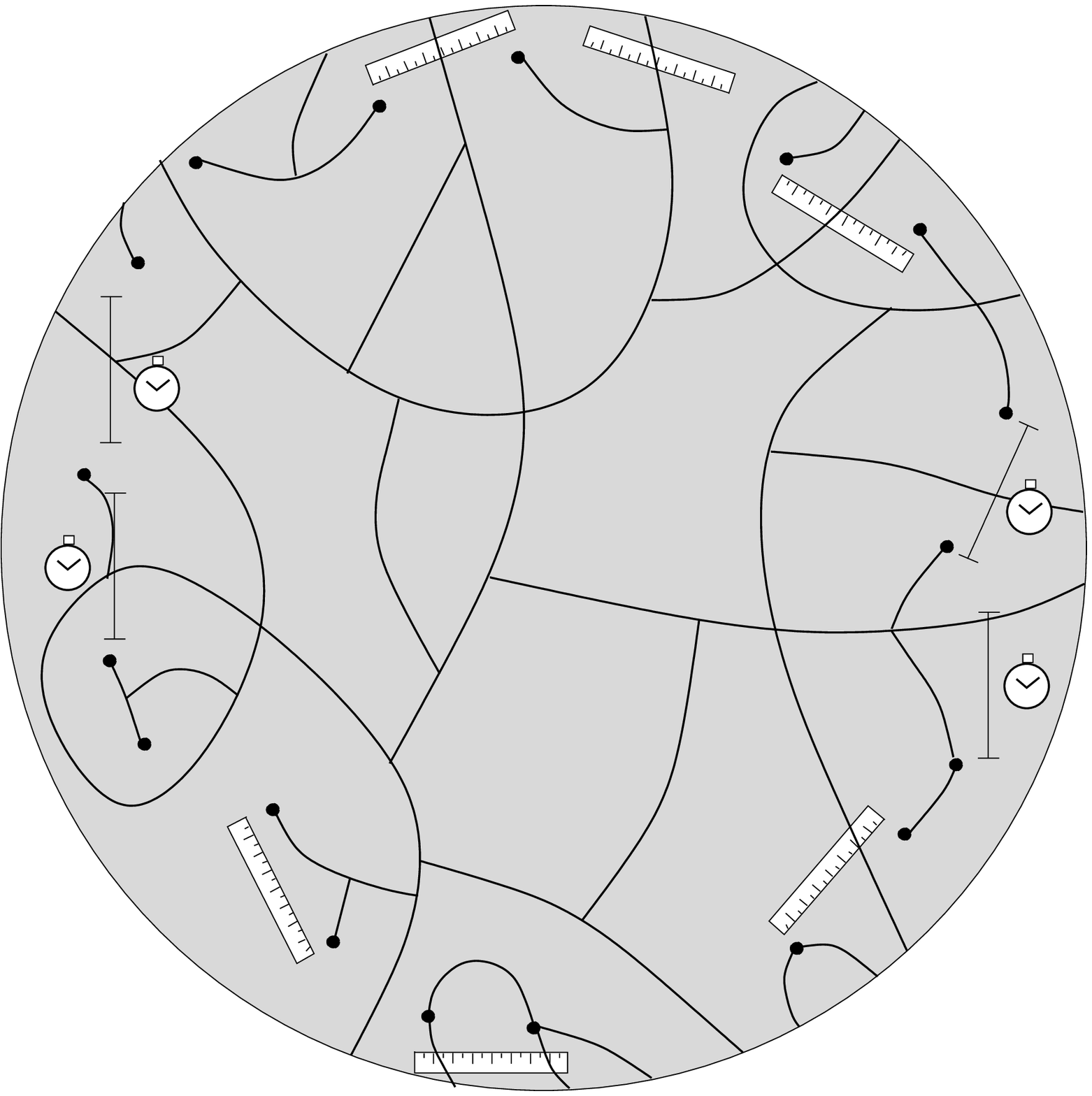} 
\end{array}
\)}
\caption{\small A system of point particles can be used as a net of Dirac observables 
to defined the physical geometry of the boundary of a local spacetime region.
In the Lorentzian sector the distance operator can have eigenvalues
corresponding to physical time. For suitable boundary states a notion of
physical time elapsed between different particles leaving the space time
region arises.}
\label{reloj}
\end{figure}
Of course there is no notion of physical time in this context as we are
dealing with Riemannian quantum gravity. In the Lorentzian context the
distance operator contains time eigenvalues \cite{Freidel}.  There is no obstruction
in principle to generalize our results to the Lorentzian sector. In that case
preliminary results indicate that a notion of Dirac time elapsed between
different events---defined in a diff-invariant way using the point
particles---can be defined in analogy to the distance operators presented
here. A representation of this idea is shown in Figure \ref{reloj}.
A systematic study of this case will be presented elsewhere.

Generalization to four dimensional gravity appears much more
technically involved for several reasons. For instance, one cannot use
point particles to define geometric operators in that case because
they create a black hole. Then, it would be necessary to consider
extended objects or matter field which complicate a lot the
procedure. All the same, this point seems very important for it could
put light on some open issues in loop quantum gravity as the meaning
of the discreteness of area operators or the interpretation of the
Immirzi-Barbero parameter.

\medskip

Finally, we would like to mention the possibility of making use of the previous results to describe
an eventual self-gravitating quantum field theory. Let us recall that a consistent quantum field
theory coupled to the gravitational field does not still exist even in
the three dimensional case. The main obstruction to define such a
theory is that we cannot use standard quantum field theory methods: if
one tries to quantize the coupling between the gravitational field and
a scalar field for example using perturbative methods (developing
around the flat metric), it encounters that the theory is
non-renormalizable. In fact, the philosophy of that method consists of
quantizing first the field in a flat spacetime and then considering
the quantum fluctuations of the gravitational field. 
Our result here provides a background independent quantization of a system of
particles. As such it could be regarded as a first quantized version of
particles and gravity coupled together providing the notion of $n$-particle
Hilbert space ${\Hp^{(N)}}$. There is no obstruction in putting together these Hilbert 
spaces in a second quantized version of the theory. 
Indeed, generalized spin-network states with $N$ particles can be taken as an
element of
\[{\cal H}=\bigoplus_{N} {\Hp^{(N)}}.\]
As Fock space is the starting point for the definition of perturbative
technique in standard quantum field theory, the previous Hilbert space could
be of interest for the definition of generally covariant quantum field
theories in three dimensions.
To define the model, it is necessary to introduce a kind of interaction
between particles that would mimic the self-interaction of a field. Therefore,
this would modify the dynamics of the theory and the physical Hilbert space
would be completely different than the one we have previously described. 
The implications of this viewpoint are currently under investigation.
\vskip1cm
\noindent {\em Avant tout... A nos Amours.}
\vskip1cm
\subsubsection*{Acknowledgments}
We want to thank Laurent Freidel and David Louapre to have shared with us some
of their results before publication. We would like also to thank Abhay
Ashtekar, Rodolfo Gambini, Michael Reisenberger, and Carlo Rovelli for 
stimulating discussions. This work has been supported by NSF
grants PHY-0090091 and INT-0307569 and the Eberly Research Funds of Penn State
University.

\newpage

\subsection*{APPENDIX: Euclidean Group $ISU(2)$}
The Euclidean group $G=ISU(2)$ is the isometry group of the three dimensional Euclidean space $\mathbb E^3$. It is a six dimensional group and therefore any of its elements is parametrized an element $z=(z^I)_I \in \mathbb R^6$; we will denote by $X(z)$ the element of $G$ parametrized by $z$.

Its Lie algebra $\mathfrak g$ is generated by infinitesimal generators of the three translations $(P_i)_{i=0,1,2}$ and infinitesimal generators of the three rotations $(J_i)_{i=0,1,2}$ which satisfy the following relations:
\begin{eqnarray}
[P_i,P_k] \; = \; 0 \;\; , \;\; [P_i,J_k] \; = \; \epsilon_{ik}{}^lP_l \;\; \text{and} \;\; [J_i,J_k] \; = \; \epsilon_{ik}{}^l J_l \;.
\end{eqnarray}
In these commutation relations, $\epsilon^{ikl}$ denotes the totally antisymmetric tensor and indices are lowered and raised by the Euclidean flat metric $\delta_{ik}=\text{diag}(+1,+1,+1)$. We will generically denote by $(\xi_I)_I$ the six dimensional basis of $\mathfrak g$. 

The Lie algebra $\mathfrak g$ is endowed with an non-degenerated invariant bilinear form (Killing form) $<,>: \mathfrak g \times \mathfrak g \rightarrow \mathbb C$ defined by:
\begin{eqnarray}\label{Killingform}
<J_i,P_k> \; = \; \delta_{ik} \;\; \text{and} \;\; <J_i,J_k> \; = \; 0 \; = \; <P_i,P_k> \;.
\end{eqnarray}
Up to a normalization factor, this non-degenerated Killing form is unique. Nevertheless, it is interesting to mention the existence of a degenerated Killing form $(,): \mathfrak g \times \mathfrak g \rightarrow \mathbb C$ defined as follows:
\begin{eqnarray}
(J_i,J_k) \; = \; \delta_{ik} \;\; \text{and} \;\; (P_i,P_k) \; = \; 0 \; = \; (P_i,J_k) \;.
\end{eqnarray}
Let us denote by $t \in \mathfrak g \otimes \mathfrak g$ the Casimir tensor associated to the non-degenerated Killing form (\ref{Killingform}), i.e. $t_{IJ}=<\xi_I,\xi_J>$; the components of its inverse are denoted $t^{IJ}$. It is also useful to introduce the dual basis $(\xi^I)_I$ defined by the relation $\xi^I = t^{IJ} \xi_J$.

\medskip

In this article, we identify the configuration space $\cal C$ of a relativistic particle with the group $G$ and Euclidean transformations on the particle are given by the left multiplication (\ref{Poincaretransformations}):
\begin{eqnarray}
G \times {\cal C} \; \longrightarrow \; {\cal C} \;\;,\;\;\;\;\; (g,X) \; \longmapsto \; gX \;. 
\end{eqnarray}
The left multiplication is completely determined by the application $f: \mathbb R^6 \times \mathbb R^6 \rightarrow \mathbb R^6$ defined as follows:
\begin{eqnarray}\label{functionf}
g(\alpha) \; X(z) \; = \; X(f(\alpha,z)) \;,
\end{eqnarray}
where we have introduced the notation $\alpha=(\alpha^I)_I\in \mathbb R^6$ and we have considered the element $g(\alpha) \in G$ defined by $g(\alpha)=\exp(-\alpha^I \xi_I)$. In the article, we make use of the infinitesimal version of (\ref{functionf}) which reads:
\begin{eqnarray}
-\xi_I \; X(z)\; = \; \frac{\partial X}{\partial z^J} \; \frac{\partial f^J}{\partial \alpha^I}(\alpha=0,z) \;.
\end{eqnarray}

\medskip

By definition, the Euclidean group $G$ is the semi-direct product of the group of three dimensional translations $I$ by the group of three dimensional rotations $SU(2)$. In the vectorial representation, this property is expressed by the fact that any element $X \in G$ is represented by a $4\times 4$ matrix as follows:
\begin{eqnarray}
X \; = \; \left( \begin{array}{cc} \Lambda & q \\ 0 & 1 \end{array} \right) \;.
\end{eqnarray}
In this representation, $\Lambda$ is a $SU(2)$ matrix evaluated in the spin 1-representation and $q$ is an element of $\mathbb R^3$. We will denote by $X=(\Lambda,q)$ this element and one can easily show that the group law is given by:
\begin{eqnarray}
(\Lambda_1,q_1) \; \cdot \; (\Lambda_2,q_2) \; = \; (\Lambda_1 \Lambda_2 \; ,\;  \Lambda_1 q_2 + q_1) \;. 
\end{eqnarray}
As a result, the inverse element of $X=(\Lambda,q)$ reads $X^{-1}=(\Lambda^{-1}, - \Lambda q)$. Given two elements $X,Y \in G$, we introduce the notation $X_1Y_2 = X \otimes Y \in G \otimes G$. To be more precise, we have $X_1 = X \otimes 1 = X^i_j E^j_i \otimes 1$ where $(E^i_j)_{ij}$ is the canonical basis of the space of $4\times 4$ matrices.

\medskip

The group $SU(2)$ is a sub-group of the Euclidean group. Unitary, irreducible representations of $SU(2)$ are finite dimensional and labeled by a spin $j \in \frac{1}{2} \mathbb N$. Given a half-integer $j$, we use the notation $\stackrel{j}{\pi}$ for the representation. The associated vector space is denoted $\stackrel{j}{V}$, it is of dimension $\Delta_j=2j+1$ and we introduce one orthonormal basis $\{\stackrel{j}{e}{}\!\!^m_n \vert m,n \in [-j,+j]\}$. These representations are also representations of the algebra $su(2)$ and the basis is chosen such that it diagonalizes the element $J_0 \in su(2)$.

\bibliographystyle{unsrt}

\begin{thebibliography}{10}

\bibitem{Ashtekar2}
A. Ashtekar, J. Lewandowski,
\newblock{``Background independent quantum gravity: a status report'',}
\newblock{In preparation.}


\bibitem{Ashtekar}
A. Ashtekar, J. Lewandowski,
\newblock{``Quantum theory of gravity I: Area operators'',}
\newblock{Class. Quant. Grav. 14, A55, (1997).}


\bibitem{Babelon}
O. Babelon,
\newblock{``Construction of the classical $r$-matrices for the Toda and Calogero models'',}
\newblock{preprint hep-th/9306102.}

\bibitem{Bojowald}
M. Bojowald,
\newblock{``Isotropic loop quantum cosmology'',}
\newblock{Class. Quant. Grav. 19, 2717-2742, (2002).}

\bibitem{Buffenoir}
E. Buffenoir, K. Noui,
\newblock{``Unfashionable observations about 3 dimensional gravity'',}
\newblock{preprint gr-qc/0305079.}

\bibitem{BNR}
E. Buffenoir, K. Noui, P. Roche,
\newblock{``Hamiltonian quantization of Chern-Simons theory with $SL(2,\mathbb C)$ group'',}
\newblock{Class. Quant. Grav. 19, 4953-5014, (2002).}

\bibitem{Carlip}
S. Carlip,
\newblock{``Quantum gravity in 2+1 dimensions'',}
\newblock{Cambridge University Press UK, 276p, (1998).}

\bibitem{carloro}
F. Condary, L. Doplicher, R. Oeckl, C. Rovelli, M. Testa,
\newblock{``Minkowxski vaccum in background independent quantum gravity'',}
\newblock{pre-print gr-qc/0307118.}

\bibitem{Deser}
S. Deser, R. Jackiw, G. t'Hooft,
\newblock{``Three dimensional Einstein gravity: dynamics of flat space'',}
\newblock{Ann. Phys. 152, 220-237, (1984).}


\bibitem{Freidel}
L. Freidel, E. Livine, C. Rovelli,
\newblock{``Spectra of Length and Area in 2+1 Lorentzian Loop Quantum Gravity'',}
\newblock{Class. Quant. Grav. 20, 1463-1478, (2003).}

\bibitem{Freidel2}
L. Freidel, D. Louapre,
\newblock{``Ponzano-Regge model revisited I: gauge fixing, observables and interacting spinning particles'',}
\newblock{preprint gr-qc/0401076}


 \bibitem{Gaul}
M. Gaul, C. Rovelli,
\newblock{``Loop quantum gravity and the meanning of diffeomorphism invariance'',}
\newblock{Lect. Notes. Phys. 541, 277-324, (2000).}


\bibitem{Goldman}
W. Goldman,
\newblock{``The symplectic nature of fundamental groups of surface'',}
\newblock{Adv. Math. 54, 200-225, (1984).}


\bibitem{Sousa}
P. de Sousa Gerbert,
\newblock{``On Spin and (Quantum) gravity in 2+1 dimensions'',}
\newblock{Nucl. Phys. B 346, 440-472, (1990).}

\bibitem{Matschull}
H.J. Matschull,
\newblock{``Quantum mechanics of a point particle in 2+1 dimensional gravity'',}
\newblock{Class. Quant. Grav. 15, 2981-3030, (1998).}

\bibitem{Noui}
K. Noui, A. Perez,
\newblock{``Three dimensional loop quantum gravity: physical scalar product and spin-foam models'',}
\newblock{gr-qc/0402110.}

\bibitem{Noui2}
K. Noui, A. Perez,
\newblock{``Distance observables in three dimensional loop quantum gravity'',}
\newblock{\it in preparation.}


\bibitem{Perez}
A. Perez,
\newblock{``Spin Foam Models for Quantum Gravity'',}
\newblock{Class. Quant. Grav. 20, R43-R104, (2003).}

\bibitem{carlobook}
C. Rovelli,
\newblock{``Quantum gravity'',}
\newblock{Cambridge University Press, {\it in press}, avalaible in www.cpt.univ-mrs.fr/~rovelli}

\bibitem{Rovelli2}
C. Rovelli, L. Smolin,
\newblock{``Discreteness of Area and Volume operators in Quantum Gravity'',}
\newblock{Nucl. Phys. B 442, 593-622; Erratum-ibid. B456, 753, (1995).}

\bibitem{Rovelli3}
C. Rovelli,
\newblock{``Black hole entropy from loop quantum gravity'',}
\newblock{Phys. Rev. Lett. 14, 3288-3291, (1996).}

\bibitem{Rovelli4}
C. Rovelli,
\newblock{``What is observable in classical and quantum gravity'',}
\newblock{Class. Quant. Grav. 8, 297-316, (1991).}

\bibitem{Rovelli5}
C. Rovelli,
\newblock{``Basis of the Ponzano-Regge-Turaev-Viro-Ooguri in the loop representation basis'',}
\newblock{Phys. Rev. D 48, 2702-2710, (1993).}



\bibitem{Teteilboim}
C. Teteilboim,
\newblock{``Quantum mechanics of the gravitational field'',}
\newblock{Phys. Rev. D 12, 3159-3179, (1982).}

\bibitem{Thiemann}
T. Thiemann,
\newblock{``Introduction to Modern Canonical Quantum General Relativity'',}
\newblock{Living Rev. Rel.}

\bibitem{phoenix}
T. Thiemann,
\newblock{``The phoenix project: master constraint program for loop quantum gravity'',}
\newblock{pre-print gr-qc/0305080.}


\bibitem{Witten}
E. Witten,
\newblock{``2+1 dimensional gravity as an exactly soluble system'',}
\newblock{Nucl. Phys. B 311, 46-78, (1988).}

\end{thebibliography}

\end{document}